\documentclass[amsmath,amssymb,11pt]{article}
\usepackage{jheppub2}
\pdfoutput=1
\usepackage{graphicx,epsfig}
\usepackage{float}
\usepackage{subfig}
\usepackage{subfloat}
\usepackage[utf8]{inputenc}
\usepackage{hyperref}
\usepackage{caption}
\usepackage{color}

\newcommand*{\affmark}[1][*]{\textsuperscript{#1}}

\newcommand{\beq}{\begin{equation}}

\newcommand{\eeq}{\end{equation}}

\title{The Stringy Origins of Galileons and their Novel Limit}

\author{Damien Easson\affmark[1],}
\emailAdd{easson@asu.edu}
\author{Tucker Manton\affmark[1],}
\emailAdd{tucker.manton@asu.edu}
\author{Maulik Parikh\affmark[1,2],}
\emailAdd{maulik.parikh@asu.edu}
\author{and Andrew Svesko\affmark[3]}
\emailAdd{a.svesko@ucl.ac.uk}

\affiliation{\affmark[1]Department of Physics, Arizona State University, Tempe, Arizona 85287, USA\\
\affmark[2]Beyond: Center for Fundamental Concepts in Science, Arizona State University,\\
 Tempe, Arizona 85287, USA\\
\affmark[3]Department of Physics and Astronomy, University College London,\\
Gower Street, London, WC1E 6BT, United Kingdom}

\abstract{We show generalized Galileons -- a particular subclass of Horndeski gravity -- arise from a consistent Kaluza-Klein reduction 
of the low-energy effective action of heterotic string theory to first order in $\alpha'$. This suggests Horndeski theories of gravity have a string-theoretic origin. The form of the Galileonic terms is precisely fixed by parameters of the embedding spacetime, so that only a specific subset of Horndeski theories is permitted by string theory. A novel limit of the model is considered by performing a dimensionful rescaling of $\alpha'$.
}

\begin{document}

\maketitle

\section{Introduction} \label{intro}

The non-renormalizability of general relativity (GR) requires that Einstein's theory be modified by the presence of higher-curvature corrections \cite{Clifton:2011jh}. Now, Einstein gravity was traditionally thought to be the unique covariant theory of pure gravity in $D=4$ spacetime dimensions with second order equations of motion of the metric, and without any ghost-like Ostrogradsky instabilities -- a consequence of Lovelock's theorem \cite{Lovelock:1971yv,Lovelock:1972vz}. Thus any $D\leq4$ dimensional modification of gravity must involve the addition of a non-trivial field,\footnote{A caveat to this is $f(R)$ gravity, a pure modified theory of gravity in $D\leq4$ dimensions. It has been shown, however, $f(R)$ gravity is physically equivalent to Brans-Dicke theory, even at the 1-loop level \cite{Teyssandier:1983zz,Maeda:1988ab,Wands:1993uu,Magnano:1993bd}.} the simplest being a scalar $\phi$. Of the scalar-tensor theories of gravity, Horndeski models \cite{Horndeski74-1}, \emph{e.g.},  covariantized Galileons
\cite{Deffayet:2009wt,Deffayet:2009mn} and their generalization \cite{Deffayet:2011gz}, is preferred among the many potential candidates as it yields second order equations of motion. The covariant \emph{generalized} Galileons\footnote{In this article, unless otherwise specified, when we refer to Galileons we mean the generalized Galileons characterized by the Lagrangian (\ref{Horndaction}).} Lagrangian (as written in, \emph{e.g.}, \cite{Kobayashi:2011nu,Charmousis14-1}) is
 \beq 
 \begin{split}
 \mathcal{L}&=K(\phi,X)-G_{3}(\phi,X)\Box\phi+G_{4}(\phi,X)R+G_{4,X}\left[(\Box\phi)^{2}-(\nabla_{\mu}\nabla_{\nu}\phi)^{2}\right]\\
 &+G_{5}(\phi,X)G_{\mu\nu}\nabla^{\mu}\nabla^{\nu}\phi-\frac{1}{6}G_{5,X}\left[(\Box\phi)^{3}-3\Box\phi(\nabla_{\mu}\nabla_{\nu}\phi)^{2}+2(\nabla_{\mu}\nabla_{\nu}\phi)^{3}\right]\;,
 \end{split}
\label{Horndaction} \eeq
where $X\equiv(\nabla\phi)^{2}$, and $K,G_{3,4,5}$ are general functions of $\phi$ and $X$. Written in this way, it is easy to recognize various limits of the four-dimensional theory (up to total derivatives), including general relativity ($K=0$, $G_{3,4,5}$ are constants), Brans-Dicke \cite{Brans:1961sx} ($G_{4}=\phi$, $K=X$, $G_{5}$ is a constant),  K-essence \cite{ArmendarizPicon:1999rj,ArmendarizPicon:2000dh,ArmendarizPicon:2000ah,Deffayet:2011gz} ($G_{3,4,5}$ are constant), and the `Fab Four' \cite{Charmousis:2011bf,Charmousis:2011ea} ($K=0$, and a combination of $G_{i}$'s which we will describe in more detail in a later section).

 Originally a local modification to general relativity\footnote{When not coupled to gravity, Galileon theories have the inerpretation of a Goldstone boson $\phi$ in flat space which is invariant under the shift symmetry $\phi(x)\to\phi(x)+b_{\mu}x^{\mu}+c$ for arbitrary parameters $b^{\mu}$ and $c$. The covariantized Galileons, and their generalizations, however, break this Galilean symmetry \cite{Deffayet:2009wt}.} \cite{Nicolis:2008in}, Galileons have proven particularly useful in describing the inflationary phase of the universe \cite{Silva:2009km,Creminelli10-1,Kobayashi:2010cm}, dark energy \cite{Deffayet:2010qz}, and bouncing cosmologies \cite{Qiu:2011cy,Easson:2011zy,Rubakov13-1,Ijjas16-1} as the generalized Galileons ``naturally'' violate the null energy condition (NEC).  For all its appeal, however, Horndeski theory lacks a fundamental origin; it is simply one of several potential scalar-tensor theories of gravity. It would be desirable to show the Galileons arise from a more fundamental theory. 

Some progress was made in understanding the fundamental origins of Galileons by recognizing they arise from higher-dimensional Lovelock theories of gravity via a standard Kaluza-Klein (KK) dimensional reduction \cite{VanAcoleyen11-1}. The result is easy to anticipate: Lovelock and Horndeski theories of gravity are both characterized by the fact that, although they include higher derivative terms in the action, their equations of motion in the metric and scalar field remain second order. In KK reduction, the higher-dimensional metric $\hat{g}_{MN}$ can be naturally decomposed into a lower-dimensional metric $g_{\mu\nu}$, a vector $A_{\mu}$, and a scalar $\phi$. A consistent reduction of a higher-dimensional pure Lovelock theory of gravity then reduces to a second-order theory of non-minimally coupled scalar-vector-gravity. The reduced theory must be second order since the lower dimensional fields are components of the higher dimensional metric solution to pure Lovelock gravity.  The scalar-tensor sector of the reduced action, as we will review, are of the form (\ref{Horndaction}). Thus, (generalized) Galileons arise from Lovelock gravity. 

Pure Lovelock gravity, however, also suffers from the absence of a fundamental origin. Moreover, since there are only a finite number of pure Lovelock terms in higher dimensions, pure Lovelock gravity is not renormalizable. Lovelock contributions do, however, appear naturally  in a more fundamental, UV complete theory, namely, superstring theory. Specifically, the Gauss-Bonnet term appears as the $\alpha'$ correction to the gravi-dilaton sector of the low-energy effective action of $10$-dimensional heterotic string theory \cite{Zwiebach85-1,Sen:1985qt,Gross86-1,Gross:1986mw,Metsaev:1987zx}. The Gauss-Bonnet term appears in order to rid the low-energy theory of ghosts and for the $\alpha'$ corrected Polyakov string action to obey conformal invariance at the quantum level.

 In the first part of this article, building on \cite{VanAcoleyen11-1,Charmousis12-1}, we present our central result: \emph{Galileon theories can be embedded in 10-dimensional string theory}. Specifically, we show  a particular subset of the generalized Galileons arises from a KK reduction of the gravi-dilaton sector of the $\alpha'$-corrected low-energy effective action of heterotic string theory.  We will see Brans-Dicke, K-essence, and the Fab four, all appear as sub-classes of the reduced theory. 

The KK reduction of a higher dimensional theory with a Gauss-Bonnet term -- either in pure gravity or in string theory -- provides a mechanism for the Gauss-Bonnet term to affect the local dynamics in four spacetime dimensions.  Recently, however, the authors of \cite{Glavan:2019inb} uncovered a way to sidestep Lovelock's theorem and write down Einstein-Gauss-Bonnet gravity (EGB) directly in $D=4$ dimensions. The novel theory arises from the simple dimensional rescaling 
\beq \alpha_{GB}\to \frac{\alpha_{GB}}{(D-4)}\;.\label{GBcouplingrescale} \eeq
Consequently, the Gauss-Bonnet term contributes to the local $D=4$ dynamics, yielding a non-trivial modification to GR in four dimensions. Applying this trick has led to a flood of papers devoted to the study of this pure covariant theory of gravity, \emph{e.g}., \cite{Glavan:2019inb,Konoplya:2020qqh,Kumar:2020owy,Mishra:2020gce,Doneva:2020ped,Odintsov:2020sqy,Jusufi:2020yus}. Nevertheless, this approach has come under some scrutiny (c.f. \cite{Arrechea:2020evj,Aoki:2020lig,Lu:2020iav,Gurses:2020ofy,Ai:2020peo,Mahapatra:2020rds}) leading to two alternative ways to take the $D=4$ limit of EGB \cite{Lu:2020iav,Fernandes:2020nbq,Hennigar:2020lsl,Easson:2020mpq}, which remarkably coincide. 

The first of these alternative novel theories \cite{Lu:2020iav} was accomplished via a KK reduction over a maximally symmetric internal space \cite{VanAcoleyen11-1,Charmousis:2014mia}, in addition to the rescaling of the Gauss-Bonnet coupling (\ref{GBcouplingrescale}). Thus, in this alternative approach, novel EGB theory is not a pure theory of gravity, but rather a special case of Horndeski gravity induced by KK reduction of pure EGB. We emphasize this novel Horndeski theory is markedly different from other Horndeski theories; namely, the trace of the gravitational equations of motion of the novel Horndeski theory is equivalent to the trace of the equations of motion of the pure novel EGB theory \cite{Glavan:2019inb}. The second alternative novel theory \cite{Fernandes:2020nbq,Hennigar:2020lsl,Easson:2020mpq} is inspired by the $D\to2$ limit of GR by  Mann and Ross \cite{Mann:1992ar}, in which one performs a type of dimensional regularization involving the subtraction of a conformally transformed pure EGB action, in addition to the rescaling of the Gauss-Bonnet coupling. Remarkably, up to simple field rescalings, the alternative novel EGB models -- KK reduction and dimensional regularization -- match identically.

In \cite{Easson:2020mpq}, moreover, a $D\to4$ limit of the low-energy effective heterotic string action was written down \emph{without} performing a dimensional reduction. In order to have a well-defined $D\to4$ limit of the $\alpha'$-corrected theory, it was demonstrated that one must also perform a dimensional rescaling of the 10-dimensional dilaton, $\phi\to (D-4)\phi$. The novel effective string thus points to an important difference between the $D\to4$ limits of pure EGB developed in \cite{Lu:2020iav} and \cite{Fernandes:2020nbq,Hennigar:2020lsl,Easson:2020mpq}. To better understand this difference, a second goal of this article is to follow the approach of \cite{Lu:2020iav} and carry out the $D=4$ limit of the (more fundamental) $\alpha'$-corrected theory. This is accomplished via a consistent KK reduction and a rescaling of $\alpha'$ akin to (\ref{GBcouplingrescale}). We will draw comparisons to this method and the novel string theory written in \cite{Easson:2020mpq}. 
 
The structure of this article is as follows. In Section \ref{HfromL} we present an explicit derivation of Horndeski gravity from the KK reduction of pure EGB over a flat $n$-dimensional internal space, following the work of \cite{VanAcoleyen11-1}. For completeness and pedagogy, we relegate many of the calculational details to Appendix \ref{appBKKhigher}. We perform a similar calculation in Section \ref{galsfromstrings} and uncover the stringy origins of the Galileons. After briefly reviewing the novel EGB model defined via dimensional regularization, Section \ref{novelST} examines whether a novel string action in $D=4$ found by KK reduction plus a rescaling of $\alpha'$ is possible. Finally, in Section \ref{conclusion}, we discuss some interesting consequences of our results, including questions about the consistency of the violation of the null energy condition and about generating the $p$-form Galileons. We also point to several directions for future research.


\section{Horndeski from Lovelock: A Review} \label{HfromL}

In $D\geq5$ dimensions, Lovelock theories of gravity generalize Einstein's general relativity to Einstein-Lovelock theories of gravity\footnote{By Einstein-Lovelock theories of gravity, we really mean to start our Lovelock Lagrangian at $n=2$. For even $D$, $t=D/2$, while for odd $D$, $t=(D-1)/2$. When $2n>D$, the quantity $\sqrt{-g}\mathcal{R}_{(n)}$ is the generalized Euler-density in $2n$-dimensions, 
$$\chi(\mathcal{M}_{2n})=\frac{1}{(4\pi)^{n}n!}\int_{\mathcal{M}_{2n}}d^{2n}x\sqrt{-g}\mathcal{R}_{(n)}\;.$$
This tells us that $\mathcal{L}_{(n)}$ is topological in $2n$-dimensional spacetimes, while for $D>2n$, $\mathcal{L}_{(n)}$ contributes to local dynamics.}
\beq I_{ELL}=\int d^{D}x\sqrt{-g}\left[\frac{1}{16\pi G_{D}}(R-2\Lambda_{0})+\mathcal{L}_{LL}\right]\;.\label{ELLactionD}\eeq
Here $\Lambda_{0}$ is the bare cosmological constant, $G_{D}$ is the $D$-dimensional Newton's constant and the Lovelock Lagrangian is 
\beq \mathcal{L}_{LL}=\sum_{n=0}^{t}\mathcal{L}_{(n)}=\sqrt{-g}\sum_{n=0}^{t}\alpha_{n}\mathcal{R}_{(n)}\;,\quad \mathcal{R}_{(n)}=\frac{1}{2^{n}}\delta^{\mu_{1}\nu_{1}...\mu_{n}\nu_{n}}_{\alpha_{1}\beta_{1}...\alpha_{n}\beta_{n}}\prod_{r=1}^{n}R^{\alpha_{r}\beta_{r}}_{\;\;\;\;\;\;\;\mu_{r}\nu_{r}}\;,\eeq
where $\alpha_{n}$ are Lovelock coupling constants and we have used the generalized Kronecker delta symbol
\beq \delta^{\mu_{1}\nu_{1}...\mu_{n}\nu_{n}}_{\alpha_{1}\beta_{1}...\alpha_{n}\beta_{n}}=n!\delta^{\mu_{1}}_{[\alpha_{1}}\delta^{\nu_{1}}_{\beta_{1}}...\delta^{\mu_{n}}_{\alpha_{n}}\delta^{\nu_{n}}_{\beta_{n}]}\;.\eeq
The first few terms are
\beq
\begin{split}
&\mathcal{L}_{(0)}=\alpha_{0}\;,\quad \mathcal{L}_{(1)}=\alpha_{1}R\;,\\
&\mathcal{L}_{(2)}\equiv \mathcal{L}_{GB}=\alpha_{GB}(R^{2}-4R_{\mu\nu}R^{\mu\nu}+R_{\mu\nu\rho\sigma}R^{\mu\nu\rho\sigma})\;.
\end{split}
\eeq
The field equations for pure Lovelock are 
\beq \mathcal{G}_{\mu\nu}=\sum_{n=0}^{t}\alpha_{n}\mathcal{G}^{(n)}_{\mu\nu}=\frac{1}{2}T_{\mu\nu}\;,\quad \mathcal{G}^{(n)\alpha}_{\;\beta}=-\frac{1}{2^{n+1}}\delta^{\alpha\mu_{1}\nu_{1}...\mu_{n}\nu_{n}}_{\beta\sigma_{1}\rho_{1}...\sigma_{n}\rho_{n}}\prod_{p=1}^{n}R^{\mu_{p}\nu_{p}}_{\;\;\;\;\;\;\;\sigma_{p}\rho_{p}}\;,\eeq
which will vanish identically for $D\leq 2n$, due to the totally antisymmetric Kronecker delta symbol. In $D=3,4$, the Einstein-Hilbert action is the only Lovelock theory which influences the dynamics non-trivially -- with the pure Gauss-Bonnet Lagrangian being a topological term, a consequence which was recently sidestepped by the dimensionful rescaling of the Gauss-Bonnet coupling (\ref{GBcouplingrescale}) \cite{Glavan:2019inb}. 

As is well known, Kaluza and Klein showed that the Einstein-Hilbert Lagrangian in $D+1$ dimensions reduces to a $D$-dimensional theory of gravity, plus Maxwell fields and a dilaton. Here we will present an explicit example of Horndeski theory arising from the dimensional reduction of Einstein-Gauss-Bonnet gravity. This calculation was accomplished for general theories of Lovelock gravity in \cite{VanAcoleyen11-1} which, however, ignored the influence of the vector potential.\footnote{The non-diagonal dimensional reduction of Einstein-Gauss-Bonnet over a circle was previously accomplished in \cite{MuellerHoissen:1989yv}. It is in fact classically consistent to set $A_{\mu}=0$, and results only in the scalar-tensor sector. Moreover, when $A_{\mu}\neq0$, it will not alter the form of the scalar-tensor sector.} Of particular interest to us is the `diagonal' reduction of the Einstein-Hilbert and Gauss-Bonnet terms over a flat $n$-dimensional Euclidean space (a $n$-torus). An expression for this quantity was previously given in \cite{Charmousis12-1}; however, in their calculation they discarded total derivative terms which we will need to know explicitly. Therefore, we carry out the calculation in full, with many details relegated to Appendix \ref{appBKKhigher}.

Following \cite{Duff86-1,Cvetic00-1}, we take as our consistent metric ansatz for reducing Einstein-Gauss-Bonnet gravity over an $n$-dimensional internal manifold to be
\beq d\hat{s}^{2}_{(p+1+n)}=e^{2\alpha\phi}ds^{2}_{p+1}+e^{2\beta\phi}dK^{2}_{(n)}\;. \label{metansatz2}\eeq
We will assume a diagonal reduction of the metric, \emph{i.e}., $A_{\mu}=0$, and $dK^{2}_{(n)}$ encodes the geometry of a flat $n$-dimensional Euclidean space. We assume the internal space is flat for the sake of simplicity,\footnote{Our $n$-dimensional Euclidean space need not be flat, as noted in \cite{Charmousis12-1}. However, the consistency condition needs to be explicitly checked. Consistency is satisfied provided the internal space being compactified is a Gauss-Bonnet space, \emph{i.e.}, a Euclidean manifold whose Ricci and Lanczos tensors are proportional to the metric. In fact, here the number of reduced dimensions $n$ can be analytically continued to real values such that the reduction generates a continuous family of theories.  For additional discussion on relevant consistent dimensional reductions, see also \cite{Gouteraux:2011qh}.} \emph{i.e.}, all higher dimensional solutions to the Einstein-Gauss-Bonnet theory respecting the symmetries of the ansatz (\ref{metansatz2}) will be solutions to the reduced theory, and, vice versa, all lower dimensional solutions can be oxidized to $p+1+n$ dimensions, likewise respecting the symmetries of the solutions to the higher dimensional theory. Lastly, here $\alpha$ and $\beta$ are constants which we will have some freedom in choosing for convenience.

Here we further assume that we need only keep the lowest modes of the KK tower, \emph{i.e.}, $g_{\mu\nu}=g_{\mu\nu}(x)$ and $\phi=\phi(x)$. This assumption is allowed since the isometry group of the internal space we compactify over is Abelian, in which case the massive modes of the KK reduction are truncated away\footnote{As argued in \cite{Charmousis12-1}, the massive modes in the KK tower transform as doublets under the action of the group while zero modes transform as singlets. When the group is Abelian, these sectors decouple.}.

Following the same procedure as described above (see Appendix \ref{appBKKhigher} for details) it is straightforward to show that the Einstein-Hilbert action becomes \cite{Charmousis12-1}:
\beq
\begin{split}
\sqrt{-\hat{g}}\hat{R}&=\sqrt{-g}e^{[(p-1)\alpha+n\beta]\phi}\biggr\{R+Y\Box\phi+Z(\nabla\phi)^{2}\biggr\}\;,
\end{split}
\label{EHterm1}\eeq
where
\beq Y\equiv-2(p\alpha+n\beta)\,,\quad Z\equiv-(p(p-1)\alpha^{2}+n(n+1)\beta^{2}+2n(p-1)\alpha\beta)\;.\eeq

Meanwhile, the Gauss-Bonnet term, after dropping some total derivatives for the time being, takes the form \cite{Charmousis12-1}
\beq \begin{split}
\sqrt{\hat{g}}\hat{\mathcal{L}}_{GB}&=\sqrt{-g}e^{[(p-3)\alpha+\beta n]\phi}\biggr\{\mathcal{L}_{GB}+B_{1}G^{ab}(\nabla_{a}\phi)(\nabla_{b}\phi)+B_{2}(\nabla\phi)^{4}+B_{3}\Box\phi(\nabla\phi)^{2}\biggr\}\;.
\end{split}
\label{Ghatscaltenhighv3}\eeq
Here
\beq
\begin{split}
&B_{1}\equiv -4\left[(p-2)(p-3)\alpha^{2}+2n(p-2)\alpha\beta+n(n-1)\beta^{2}\right]\;,\\
&B_{2}\equiv-2\biggr[(p-1)(p-2)(p-3)\alpha^{3}\\
&+3n(p-2)(p-1)\alpha^{2}\beta+3n(p-1)(n-1)\alpha\beta^{2}+n(n-1)(n-2)\beta^{3}\biggr]\;,\\
&B_{3}\equiv-(p-1)(p-2)^{2}(p-3)\alpha^{4}-4n(p-1)(p-2)^{2}\alpha^{3}\beta\\
&-2n(p-1)[3pn-2p-5n+3]\alpha^{2}\beta^{2}-4n(n-1)^{2}(p-1)\alpha\beta^{3}-n(n-1)^{2}(n-2)\beta^{4}\;.
\end{split}
\eeq

Thus, the reduced Gauss-Bonnet action exactly reproduce the  Horndeski theories of gravity given by action (\ref{Horndaction}), however, with specific choices of the coefficients, and an overall conformal factor. Specifically, we have exchanged $G^{ab}\nabla_{a}\nabla_{b}\phi$ for a $G^{ab}(\nabla_{a}\phi)(\nabla_{b}\phi)$ term, up to total derivatives, and similarly for $[(\nabla_{a}\nabla_{b}\phi)^{2}-(\Box\phi)^{2}]$. Note that the reduction of the Gauss-Bonnet term alone does not yield the six-derivative quantity $G_{5,X}$ in the Galileon action (\ref{Horndaction}). Such a term may be generated by KK reduction of Lovelock terms beyond Gauss-Bonnet \cite{VanAcoleyen11-1,Charmousis:2011bf}. Moreover, due to the presence of the conformal factor, generally the KK reduced Horndeski theory lacks a constant shift symmetry in $\phi$.

We emphasize the coefficients of the Horndeski terms are fixed by the parameters $\alpha,\beta$, and the dimensions of the uncompactified and compactified spacetimes, $p+1$ and $n$, respectively. Notice when we set $\alpha=0$ the coefficients $B_{i}$ simplify considerably. Incidentally, $\alpha=0$, $\beta=\frac{1}{2}$ corresponds to what is called the `dual frame' in studies of holography, where the holographic dictionary for non-conformal branes is most easily set-up \cite{Kanitscheider:2009as,Gouteraux:2011qh}.

In fact, the parameters $\alpha,\beta$ are particularly important as they lead to various `frames'\footnote{We should be a little cautious in using the phrase `frames'. Naively, the combination of (\ref{EHterm1}) and (\ref{Ghatscaltenhighv3}) is a scalar-tensor theory written in \emph{string frame}, that is, the Einstein-Hilbert term is non-minimally coupled to the scalar field $\phi$ via a conformal factor. An important property of string frame, however, is that the \emph{matter} action is universally coupled to the string frame metric.  Had we included a matter action in the higher dimensional parent theory, a dimensional reduction would result in a lower dimensional action whose matter contributions are no longer universally coupled to the string frame metric. Moreover, the \emph{Einstein frame} is specifically the parameterization which is found by conformally transforming the string frame metric, and insisting the scalar field enters as a canonically normalized and minimally coupled scalar field. Thus, while we may choose the constants $\alpha$ and $\beta$ to place our action in a convenient form, we are not necessarily, by this more precise terminology, selecting the string frame or Einstein frame.} of interest. For example, the `Einstein frame', where the Einstein-Hilbert term (\ref{EHterm1}) is put into a canonical form:
\beq \sqrt{-g}(R-\frac{1}{2}(\nabla\phi)^{2}+...)\;,\eeq
is arrived at by the specific choice
\beq (p-1)\alpha+n\beta=0\;,\quad -Y[(p-1)\alpha+n\beta]+Z=-\frac{1}{2}\;,\eeq
where we have included the influence of the $\Box \phi$ contribution by performing an integration by parts. The Einstein frame is particularly relevant when studying the energy conditions of the theory as the resulting equations of motion clearly separate out the Einstein tensor $G_{ab}$ from other contributions.

Similarly, the `Gauss-Bonnet frame' is found by eliminating the overall dilaton factor out front, $e^{[(p-3)\alpha+\beta n]\phi}$, such that $[(p-3)\alpha+\beta n]=0$. Importantly, notice the Gauss-Bonnet and Einstein frames never coincide. As we describe in Section \ref{conclusion}, this suggests implications about the consistency of energy-conditions in lower dimensions.

Galileons emerging from Gauss-Bonnet gravity via dimensional reduction should not come as a surprise. Since the higher dimensional parent theory admits second-order equations of motion, the lower dimensional theory is guaranteed to have second order equations of motion. Crucially, moreover, while a pure Gauss-Bonnet term in $p+1\leq 4$ is purely topological, here $\mathcal{L}_{GB}$ will have an effect on lower dimensional spacetime dynamics as it couples to the scalar $\phi$ in a non-trivial way. That the Einstein-Gauss-Bonnet term influences spacetime dynamics in four dimensions is the basis of the counter proposal \cite{Liu:2020yhu} to the novel theory of Gauss-Bonnet gravity presented in \cite{Glavan:2019inb}. We will say more about this in Section \ref{novelST}, where we also discuss its string theoretic counterpart.



\section{Galileons from Strings} \label{galsfromstrings}
\indent

Above we reviewed an explicit example illuminating the purely gravitational origin of Horndeski theories of gravity -- they naturally emerge from the KK reduction of EGB. Pure theories of EGB, and Lovelock more generally, however, are not known to exist in nature, and carry renormalizability challenges of their own. The Gauss-Bonnet contribution, though, does famously arise in heterotic string theory by insisting the theory cannot have ghosts \cite{Zwiebach85-1,Gross86-1,Gross:1986mw}. More generally, Gauss-Bonnet arises by demanding the first $\alpha'$ perturbative correction to the Polyakov string action be conformally invariant at the quantum level\footnote{At $\alpha'=0$, imposing conformal invariance at the quantum level, \emph{i.e}., that the beta function vanish, requires the $D$-dimensional background spacetime the string propagates in to have vanishing Ricci curvature, $R_{ab}=0$, Einstein's vacuum equations. Equivalently, the beta function arises from a $D$-dimensional low energy \emph{effective action} (c.f. \cite{Gasperini07-1} for a pedagogical review).} \cite{Metsaev:1987zx}.  In fact, somewhat generically, the gravi-dilaton sector\footnote{We emphasize we are focusing on the gravi-dilaton sector of the string action. There of course additional sectors one could, and for total consistency, should consider, such as the Kalb-Ramond field and its field strength. Such terms, however, do not generate the \emph{scalar} Galileons we are interested in here. Rather, it is expected dimensional reduction of such sectors produce the $p$-form Galileons \cite{Deffayet:2010zh}.} of the low energy effective string action (naturally in string frame) to first order in $\alpha'$ can be written as:\footnote{For a tour de force derivation of this action, see \cite{Metsaev:1987zx}. A succinct, but more pedagogical derivation is presented in \cite{Gasperini07-1} -- with an important caveat. In \cite{Metsaev:1987zx} there is an additional coefficient multiplying the $(\hat{\nabla}\Phi)^{4}$ term (see their Eq. (3.22)), $\bar{\rho}_{1}=16\frac{(D-4)}{(D-3)^{3}}$. This term would vanish when we dimensionally reduce to $D=4$ dimensions, and would lead to a divergence in the $D=2$ limit. More generically, \cite{Gasperini07-1} shows the coefficient $\bar{\rho}_{1}$ can be dropped. Here we will follow \cite{Gasperini07-1}, however, note along the way how the physics changes if we opt for the coefficient $\bar{\rho}_{1}$ in \cite{Metsaev:1987zx}.}
\beq I=-\frac{1}{2\lambda^{D-2}_{s}}\int d^{D}x\sqrt{-\hat{g}}e^{-\Phi}\left[\hat{R}+(\hat{\nabla}\Phi)^{2}-\frac{\alpha'}{4}\hat{\mathcal{L}}_{GB}+\frac{\alpha'}{4}(\hat{\nabla}\Phi)^{4}\right]\;.\label{stringact1}\eeq
Here we have used the same hatted notation used in the previous section to denote $D$-dimensional spacetime quantities, in anticipation of dimensionally reducing this action.\footnote{We should also point out that writing the action (\ref{stringact1}) comes with an inherent field redefinition ambiguity; indeed the form of (\ref{stringact1}) comes from a specific choice in field redefinition. In light of this, we note that another field redefinition allows one to write down an action which already includes many of the Galileon terms at the level of $D$-dimensional action (see, for example, Eqs. (2.1) and (2.3) of \cite{Maeda:2011zn}. Dimensionally reducing such an action will of course also result in Galileons in lower dimensions. Consequently, due to the ambiguity in the field redefinition, the coefficients $a_{i}$ appearing in the reduced action may not in fact be unique.}

Let's now use our previous metric ansatz (\ref{metansatz2}) for compactifying a $D$-dimensional spacetime over $n$-internal dimensions. Here we assume that the dilaton $\Phi$ depends only on the lower dimensional coordinates, not on the coordinates of the $n$-dimensional internal space, \emph{i.e.}, $\Phi\equiv\phi(x)$. This is accomplished by noting that it is perfectly consistent to truncate the higher modes in the Kaluza-Klein tower, that is,
\beq \Phi(x,y)=\sum_{k=0}\Phi_{k}(x)e^{2\pi i ky}\approx\Phi_{0}(x)\equiv\phi(x)\;.\eeq
This means 
\beq (\hat{\nabla}\Phi)^{2}=\hat{g}^{MN}\hat{\nabla}_{M}\Phi\hat{\nabla}_{N}\Phi=\hat{g}^{MN}\partial_{M}\Phi\partial_{N}\Phi=e^{-2\alpha\phi}g^{ab}\partial_{a}\phi\partial_{b}\phi=e^{-2\alpha\phi}(\nabla\phi)^{2}\;,
\eeq
and, likewise,
\beq (\hat{\nabla}\Phi)^{4}=e^{-4\alpha\phi}(\nabla\phi)^{4}\;.\eeq

Altogether then, assuming a flat internal Euclidean space, the gravi-dilaton sector of the first $\alpha'$-corrected $p+1+n$-dimensional effective string action (\ref{stringact1}) reduces to
\beq 
\begin{split}
 I&=-\frac{1}{2\lambda^{p-1}_{s}}\int d^{p+1}x\sqrt{-g}\biggr\{e^{[(p-1)\alpha+n\beta-1]\phi}\left[R+a_{1}(\nabla\phi)^{2}\right]\\
&-\frac{\alpha'}{4}e^{[(p-3)\alpha+n\beta-1]\phi}\left[\mathcal{L}_{GB}+a_{2}G^{ab}(\nabla_{a}\phi)(\nabla_{b}\phi)+a_{3}(\nabla\phi)^{2}\Box\phi+a_{4}(\nabla\phi)^{4}\right]\biggr\}\;,
\end{split} 
\label{redstringact}\eeq
where we have integrated out the internal space into the adjusted parameter $\lambda^{p}_{s}$. Here we have accounted for the total derivatives appearing in (\ref{totderivs}), whereby the coefficients\footnote{Had we chosen to use the effective string action in \cite{Metsaev:1987zx}, the only change presents itself in coefficient $a_{4}$, where the leading $-1$ is replaced with $-\frac{16(p-3+n)}{(p-1+n)^{3}}$. This will not affect the physics of the $p=2,3$ limits described below, however, it would suggest the $p=1$ limit leads to a divergence.} $a_{i}$ are:
\beq\label{ais}
\begin{split}
&a_{1}\equiv 1+p^{2}\alpha^{2}+n\beta(-2+(n-1)\beta)-p\alpha(2+\alpha-2n\beta) \;,\\
&a_{2}\equiv -4(p-3)(p-2)\alpha^{2}+4n\beta(2-(n-1)\beta)-8\alpha(p-2)(n\beta-1)\;,\\
&a_{3}\equiv-2(p-1)(p-2)(p-3)\alpha^{3}-2n(n-1)\beta^{2}[-3+(n-2)\beta]\\
&-6n(p-1)\alpha\beta[-2+(n-1)\beta]-6(p-1)(p-2)\alpha^{2}(n\beta-1)\;,\\
&a_{4}\equiv-1-(p-1)(p-2)[2+(p-2)\alpha(-4+(p-3)\alpha)]\alpha^{2}-4n(n-1)^{2}[-1+(p-1)\alpha]\beta^{3}\\
&-4n(p-1)\alpha\beta[1+\alpha(5-3p+(p-2)^{2}\alpha)]-n(n-1)^{2}(n-2)\beta^{4}\\
&-2n[-1+n-2(3n-2)(p-1)\alpha+(p-1)(3-2p+n(3p-5))\alpha^{2}]\beta^{2}\;.
\end{split}
\eeq
The reduced action (\ref{redstringact}) is the Horndeski theory embedded in the higher dimensional superstring effective action.\footnote{As in the case with dimensional reduction of pure EGB, we see our theory does not include the six derivative quantity $G_{5,X}$. It is expected, however, such a term may be recovered from the $\alpha'^{2}$ correction to the effective string action, as this correction is thought to include higher curvature Lovelock contributions.} Note our string inspired Horndeski model does not possess the shift symmetry $\phi\to\phi+const$ because of the presence of the overall $e^{\phi}$. Due to this lack in shift symmetry, it is possible to evade the `no hair theorem' for Horndeski gravity \cite{Hui:2012qt,Babichev:2016rlq}, and we expect there will exist hairy black hole and star solutions. 

It is worthwhile to spend some time comparing the reduced action (\ref{redstringact}) to the general form of the Horndeski Lagrangian in (\ref{Horndaction}). We see, up to proportionality factors  and total derivatives that are unimportant in the current context,
\beq 
\begin{split}
&K(\phi,X)=e^{A\phi}X+\alpha'e^{B\phi}X^{2}\;,\\
&G_{3}(\phi,X)=e^{B\phi}X\;,\\
&G_{4}(\phi,X)=e^{A\phi}\;,\\
&G_{5}(\phi,X)=\text{constant}\;,
\end{split}
\eeq
where recall $X=(\nabla\phi)^{2}$. Notice in our case the functions $K$ and $G_{3,4,5}$ factorize.

We should clarify we can adjust the functions $G_{3,4,5}$ by using particularly useful identities and removing total derivatives (see Eqs. (\ref{Gaussbonnredint}), (\ref{useid1}), and (\ref{useid2})). Therefore, only specific choices of $\alpha,\beta$ lead to well-known scalar tensor theories, including Brans-Dicke and K-essence \cite{ArmendarizPicon:1999rj,ArmendarizPicon:2000dh,ArmendarizPicon:2000ah}.

A particularly interesting scalar-tensor theory is the `Fab Four', given by the following Lagrangian contributions \cite{Charmousis14-1},
\beq 
\begin{split}
&\mathcal{L}_{John}=\sqrt{-g}V_{John}(\phi)G^{ab}\nabla_{a}\phi\nabla_{b}\phi\;,\quad \mathcal{L}_{Paul}=\sqrt{-g}V_{Paul}(\phi)P^{abcd}\nabla_{a}\phi\nabla_{c}\phi(\nabla_{b}\nabla_{d}\phi)\;,\\
&\mathcal{L}_{George}=\sqrt{-g}V_{George}(\phi)R\;,\quad \mathcal{L}_{Ringo}=V_{Ringo}(\phi)\mathcal{L}_{GB}\;,
\end{split}
\eeq
where $P_{abcd}=-\frac{1}{4}\epsilon^{abls}\epsilon^{cdef}R_{lsef}$ is the double dual of the Riemann tensor. The Fab Four model is novel in that it has self-tuning properties \cite{Charmousis:2011bf}\footnote{Finding maximally symmetric vacua endowed with a bulk cosmological constant which is not fixed by any of the other coupling constants appearing in the gravitational action, and is aborbed by the dynamics due to a non-trivial scalar field. Self-tuning partially solves the cosmological constant problem, however, notably, does not deal with the radiative instability of the cosmological constant.}. Specifically, self-tuning solutions exist for any of these potentials provided $V_{John}\neq0$, or $V_{Paul}\neq0$, or $V_{George}$ is not constant; $V_{Ringo}$ does not self-tune but also does not prevent any self-tuning. Moreover, as dictated by Weinberg's no-go theorem \cite{Weinberg:1988cp}, the vacuum solution is not Poincar\'e invariant, such that the scalar field will depend non-trivially on the time coordinate in a flat cosmological background.

Comparing the Fab Four to (\ref{Horndaction}), we see they arise from specific choices of the $G_{3,4,5}$ functions, and $K=0$. Meanwhile, from the reduced action (\ref{redstringact}) we clearly see we have $\mathcal{L}_{John}$, $\mathcal{L}_{George}$, and $\mathcal{L}_{Ringo}$, with
\beq V_{John}=-\frac{\alpha'}{4}e^{[(p-3)\alpha+n\beta-1]\phi}a_{2}\;,\quad V_{Ringo}=\frac{1}{a_{2}}V_{John}\;,\eeq
and 
\beq V_{George}=e^{[(p-1)\alpha+n\beta-1]\phi}\;.\eeq
Naively, it seems $V_{Paul}=0$, however, we could imagine performing field redefinitions and conformal transformations to (\ref{redstringact}) to recover this term. Importantly, we see our reduced string action includes the Fab Four and can therefore describe a self-tuning theory. 

Another reason the Fab Four is of interest is that, due to the non-trivial derivative interactions of John and/or Paul, it might lead to Vainshtein effects \cite{Vainshtein:1972sx,Babichev:2013usa}, or k-mouflage (\emph{c.f.} \cite{Babichev:2009ee}). These effects are important for solar system tests of validity as the Vainshtein mechanism applied to non-GR theories leads to the recovery of GR around massive bodies by hiding extra degrees of freedom by strong kinetic self-coupling. Since our Horndeski model includes these non-trivial interactions, it is expected to exhibit Vainshtein effects as well.

\vspace{2mm}

\noindent \textbf{Taxonomy of Frames}

\vspace{2mm}

 Let us further categorize the various frames mentioned earlier.  We now have:
\beq
\begin{split}
 &\textbf{Einstein:}\quad [(p-1)\alpha+n\beta-1]=0\;,\quad a_{1}=-\frac{1}{2}\;,\\
&\textbf{Gauss-Bonnet:}\quad  [(p-3)\alpha+n\beta-1]=0\;,\\
&\textbf{Dual:}\quad \alpha=0\;,\quad \beta=\frac{1}{2}\;.
\end{split}
\eeq
We see immediately that in the Gauss-Bonnet frame, the Gauss-Bonnet term is no longer coupled to the dilaton. Thus, in $p\leq3$, the Gauss-Bonnet term in this frame is purely topological and does not affect the local dynamics of the theory.

In the dual frame $\alpha=0$, the coefficients $a_{i}$ simplify greatly, where 
\beq 
\begin{split}
&a_{1}\to\frac{1}{4}(n-1)(n-4)\;,\quad a_{2}\to-n(n-5)\;,\\
&a_{3}\to-\frac{1}{4}n(n-1)(n-8)\;,\quad a_{4}=-1-\frac{1}{16}n(n-1)(n-2)(n-9)\;,
\end{split}
\eeq
such that each coefficient is dependent on the dimension $n$ of the internal space. If we compactify over a circle ($n=1$), we lose the $(\nabla\phi)^{2}$ and $\Box\phi(\nabla\phi)^{2}$ terms. 

Of particular interest is the Einstein frame,  which leads to the conditions
\beq\beta=\frac{1}{n}[1-(p-1)\alpha]\;,\quad -\frac{1}{n}+\frac{\alpha(p-1)}{n}\left[2-(p-1+n)\alpha\right]=-\frac{1}{2}\;,\label{einframecond}\eeq
which gives two solutions\footnote{There are also two choices for $\alpha$ and $\beta$ even in dimensionally reducing pure Einstein gravity. This is a consequence of the $\phi\to-\phi$ and $\alpha\to-\alpha$ symmetry of the metric ansatz. This symmetry is broken when by coupling to a lower dimensional gauge field, however, gives a generalization of EM duality in four dimensions when the Maxwell curvature $F^{2}$ couples to the scalar. We thank Blaise Gout\'eraux for bringing this out to us.} for $\alpha$, labeled here as $\alpha_{\pm}$, and, correspondingly, the two values $\beta_{\pm}$. We also  comment that we would have arrived to the same choice of $\beta_{\pm}$ had we instead started from the reduced action and performed a conformal transformation on the $p+1$-dimensional metric, $g_{ab}=\bar{g}_{ab}e^{2\psi}$, before selecting $\alpha,\beta$. To see this, use the standard transform for the Ricci scalar, 
\beq R=e^{-2\psi}\left[\bar{R}-2(d-1)\bar{\Box}\psi-(d-1)(d-2)(\bar{\nabla}\psi)^{2}\right]\;,\eeq
such that the tree-level action becomes
\beq \sqrt{-\bar{g}}e^{[(p-1)\alpha+n\beta-1]\phi+(p-1)\psi}\left(\bar{R}-2p\bar{\Box}\psi-p(p-1)(\bar{\nabla}\psi)^{2}+a_{1}(\bar{\nabla}\phi)^{2}\right)\;.\eeq
The conformal factor on the tree-level contribution is removed by setting $\psi=-\frac{1}{(p-1)}[(p-1)\alpha+n\beta-1]$. If we further impose the coefficient in front of $(\tilde{\nabla}\phi)^{2}$ be $-1/2$, we find precisely
\beq \frac{1}{(p-1)}[n\beta(2-(p+n-1)\beta)-1]\;,\eeq
which is equivalent to the substituting the first relation in (\ref{einframecond}) for $\alpha$.

 While we can in principle work in arbitrary dimensions, the string action (\ref{redstringact}) naturally lives in $D=p+1+n=10$ spacetime dimensions. Primarily we are interested in the case when $p=3$, such that the internal compactified space is six dimensional, $n=6$. Then, 
\beq (\alpha_{+},\beta_{+})=(\frac{1}{2},0)\,,\quad (\alpha_{-},\beta_{-})=(-\frac{1}{4},\frac{1}{4})\;.\eeq
With these choices of $\alpha_{\pm},\beta_{\pm}$, $p$, and $n$, we fix the coefficients of the Galileon contributions appearing in the dimensionally reduced string action. Specifically, we are led to the following two actions:
\beq
\begin{split}
I_{+}&\equiv-\frac{1}{2\lambda^{2}_{s}}\int d^{4}x\sqrt{-g}\biggr[\left(R-\frac{1}{2}(\nabla\phi)^{2}\right)\\
&-\frac{\alpha'}{4}e^{-\phi}\left(\mathcal{L}_{GB}+4G^{ab}(\nabla_{a}\phi)(\nabla_{b}\phi)+3\Box\phi(\nabla\phi)^{2}-(\nabla\phi)^{4}\right)\biggr]
\end{split}
\label{Iplusp3}
\eeq
\beq
\begin{split}
I_{-}&\equiv-\frac{1}{2\lambda^{2}_{s}}\int d^{4}x\sqrt{-g}\biggr[\left(R-\frac{1}{2}(\nabla\phi)^{2}\right)\\
&-\frac{\alpha'}{4}e^{\phi/2}\left(\mathcal{L}_{GB}+\frac{11}{2}G^{ab}(\nabla_{a}\phi)(\nabla_{b}\phi)+\frac{15}{4}(\nabla\phi)^{2}\Box\phi-\frac{19}{16}(\nabla\phi)^{4}\right)\biggr]\;.
\end{split}
\label{Iminusp3}\eeq
 Given our starting point, actions $I_{+}$ and $I_{-}$ are the Horndeski theories which are naturally embedded in superstring theory. Thus, Galileons have a fundamental, string theoretic origin.

We emphasize the actions $I_{+}$ and $I_{-}$ arise from the freedom in choosing the constants $\alpha$ and $\beta$ at the level of the metric ansatz (\ref{metansatz2}), where $\alpha$ satisfies a quadratic equation (\ref{einframecond}). Notably, we see $I_{+}$ cannot be transformed into $I_{-}$ via a conformal transformation of the metric or field redefinitions. From the perspective of the $p+1$-dimensional theory then, $I_{+}$ and $I_{-}$ represent genuinely different theories. Yet, the daughter actions $I_{\pm}$ follow from the same parent theory. Thus, it appears there is a hidden `duality' between $I_{+}$ and $I_{-}$ which is obscure in lower dimensions but apparent in higher dimensions.

\vspace{2mm}

\noindent \textbf{Reducing to Lower Dimensions}

\vspace{2mm}

Before moving on, note that we can in principle compactify to lower dimensional actions, however, we might not always be able to move to the Einstein frame. For example, when $p=2,n=7$, leading to a $2+1$-dimensional Horndeski theory,  we have an Einstein frame for irrational values of $\alpha$ and $\beta$, namely, 
\beq (\alpha_{+},\beta_{+})=\left(\frac{1}{8}(1+\sqrt{21})\,,\,\frac{1}{56}(7-\sqrt{21})\right)\;,\quad (\alpha_{-},\beta_{-})=\left(\frac{1}{8}(1-\sqrt{21})\,,\,\frac{1}{56}(7+\sqrt{21})\right)\;.\eeq
We thus again have two actions in Einstein frame with the following coefficients:
\beq
\begin{split}
&I_{\pm}:\; a_{2}=\frac{1}{4}(13\mp\sqrt{21})\;,\quad a_{3}=\mp\frac{9}{56}(\mp21+\sqrt{21})\;,\quad a_{4}=\frac{1}{224}(-215\pm9\sqrt{21})\;,\\
 & [(p-3)\alpha+n\beta-1]=\frac{1}{4}(-1\mp\sqrt{21})\;.
\end{split}
\eeq
Thus, it is possible to embed a $2+1$-dimensional Horndeski theory in string theory, where the Gauss-Bonnet contribution non-trivially effects the dynamics of the low dimensional theory, and can be put in Einstein frame.\footnote{Note that in $2+1$-dimensions the Gauss-Bonnet term reduces to $\mathcal{L}_{\text{GB}}=4R^{2}$.}

Alternatively, however, if we were to try to compactify further to a $1+1$-dimensional scalar-theory of gravity, the condition (\ref{einframecond}) cannot be satisfied, such that an Einstein frame does not exist. This is not so surprising as generally $1+1$ dilaton theories of gravity do not have an Einstein frame.  Two-dimensional dilaton theories of gravity will nonetheless exist in other frames. For example, when we impose $\alpha=-\frac{1}{32}$,  $\beta=-\frac{1}{8}$, we have an $\alpha'$-corrected Callan-Giddings-Harvey-Strominger (CGHS)-like model\footnote{Notably we have not included the `charge' $4\lambda^{2}$ coming from the KK reduction of the charged black hole system.} \cite{Callan:1992rs}:
\beq 
\begin{split}
I_{\text{CGHS}\alpha'}&=-\frac{1}{2}\int d^{2}x\sqrt{-g}\biggr[e^{-2\phi}\left(R+4(\nabla\phi)^{2}\right)-\frac{\alpha'}{4}e^{-\frac{31}{16}\phi}\left(+\frac{105}{16}\Box\phi(\nabla\phi)^{2}-\frac{1635}{256}(\nabla\phi)^{4}\right)\biggr]\;.
\end{split}
\eeq
where we used $\mathcal{L}_{\text{GB}}=0$ in two dimensions (using $R_{abcd}=\frac{R}{2}(g_{ac}g_{bd}-g_{ad}g_{bc})$ in $D=2$). Interestingly, the Gauss-Bonnet term has no effect in the 2D case, \emph{i.e.}, there is no $R^{2}$ contribution as one might have expected. 




\section{Novel $\alpha'$-Corrected String Theory?} \label{novelST}
\indent

Novel Einstein Gauss-Bonnet gravity \cite{Glavan:2019inb} is a pure theory of  Einstein-Gauss-Bonnet gravity where the Gauss-Bonnet correction influences the local dynamics in $D=4$ spacetime dimensions. This is accomplished by rescaling the Gauss-Bonnet coupling $\alpha_{GB}\to\alpha_{GB}/(D-4)$ at the level of the equations of motion. A number of issues have been raised to suggest novel EGB is inconsistent; upon the rescaling, the $D\to4$ limit of the action is singular \cite{Arrechea:2020evj} as well as a diverging black hole entropy computed using the Wald entropy functional \cite{Lu:2020iav}. A Hamiltonian analysis further revealed that for a consistent theory to exist with only two dynamical degrees of freedom, temporal diffeomorphism invariance must be broken \cite{Aoki:2020lig}. 

Collectively, these issues led \cite{Lu:2020iav} to propose an alternative way to take the $D=4$ limit of EGB. This is accomplished via a Kaluza-Klein dimensional reduction over a maximally symmetric internal space, \emph{viz.} \cite{VanAcoleyen11-1,Charmousis:2014mia}, in addition to the coupling rescaling. Working in the `dual frame' (specifically with $\alpha=0$ and $\beta=1$), the dimensionally reduced Einstein-Gauss-Bonnet action has an overall conformal factor $e^{n\beta\phi}$. Writing $n=D-(p+1)$, expanding to linear order in $O(D-(p+1))$, and subtracting
\beq -\frac{\alpha_{GB}}{16\pi G_{p+1}}\int d^{p+1}x\sqrt{-g}\mathcal{L}_{GB}\;,\label{gbsub}\eeq
one arrives to the following $p+1$-dimensional action
\beq
\begin{split}
 I&=\int d^{p+1}x\sqrt{-g}\biggr[\frac{1}{16\pi G_{4}}R+\alpha_{GB}(\Phi \mathcal{L}_{GB}+4G^{ab}(\nabla_{a}\Phi)(\nabla_{b}\Phi)-4(\nabla\Phi)^{2}\Box\Phi+2(\nabla\Phi)^{4})\biggr]\;,
\end{split}
\label{KKredact}
\eeq
where $\alpha_{GB}$ has been rescaled $\alpha_{GB}\to\frac{\alpha_{GB}}{(D-(p+1))}$ and the limit $D\to p+1$ has been taken.

The subtraction scheme (\ref{gbsub}) is very much in the spirit of \cite{Mann:1992ar}, which uncovered a $D\to2$ limit of general relativity\footnote{The resulting action describes a dilaton theory of gravity whose equations of motion are $R=T$. Interestingly, as noted in \cite{Rosso:2020zkk}, the scalar-tensor theory can be rewritten as a pure theory of $f(R)$ gravity.} via a type of `dimensional regularization'. In fact, as shown in \cite{Fernandes:2020nbq,Hennigar:2020lsl,Easson:2020mpq}, starting with a pure EGB theory and subtracting $\int d^{D}x\sqrt{-\tilde{g}}\alpha_{GB}\tilde{\mathcal{L}}_{GB}$, where $\tilde{g}_{\mu\nu}$ is conformally related to $g_{\mu\nu}$ via $\tilde{g}_{\mu\nu}=e^{\Phi}g_{\mu\nu}$, leads to precisely the same action (\ref{KKredact}), upon elementary rescalings of $\Phi, g_{\mu\nu}$ and $\alpha_{GB}$. In $p+1=4$, the resulting equations of motion reveal the Gauss-Bonnet contribution alone will modify the local four-dimensional dynamics. The authors of \cite{Easson:2020mpq} went further by formally establishing a novel limit of general Einstein-Lovelock gravity via the same dimensional regularization and a judicious relabeling of the Lovelock coupling, where Lovelock contributions are likewise expected to influence lower dimensional spacetimes in ways they ordinarily do not.

As emphasized in the introduction, pure theories of Lovelock gravity lack a fundamental origin, however, they do appear naturally in the $\alpha'$ corrections to the low energy effective actions of superstring theory. With this perspective in mind, here we will follow \cite{Lu:2020iav} and find a novel string theory by rescaling $\alpha'\to\alpha'/(D-(p+1))$ upon a KK reduction, and then take the $D\to p+1$ limit. It should be noted that the authors of \cite{Easson:2020mpq} attempted to write down a novel limit of a low energy effective string action using the Ross-Mann method of dimensional regularization. However, it was found the overall dilaton factor appearing in the effective string action spoils the rescaling of $\alpha'$ such that the resulting $D=4$ action is divergent. To fix this issue, a further dimensional rescaling of $\phi$ was needed. It would therefore be interesting to see whether the novel scheme used in \cite{Lu:2020iav} will lead to a different result, such that we arrive to a novel EGB theory with a fundamental stringy origin. 

Therefore, take the dimensionally reduced action (\ref{redstringact}) and set $\alpha=0$, $\beta=1$ and $n=(D-(p+1))$, such that
\beq 
\begin{split}
 I&=-\frac{1}{2\lambda^{p-1}_{s}}\int d^{p+1}x\sqrt{-g}e^{(D-(p+2))\phi}\biggr(R+a_{1}(\nabla\phi)^{2}\\
&-\frac{\alpha'}{4}\left[\mathcal{L}_{GB}+a_{2}G^{ab}(\nabla_{a}\phi)(\nabla_{b}\phi)+a_{3}(\nabla\phi)^{2}\Box\phi+a_{4}(\nabla\phi)^{4}\right]\biggr)\;,
\end{split} 
\label{novelstv1}\eeq
where now\footnote{Where again, if we use the effective action in \cite{Metsaev:1987zx}, the final $-1$ in coefficient $a_{4}$ becomes proportional to $(p-3)/(p-1)^{3}$.}, 
\beq
\begin{split}
&a_{1}=(D-(p+1))(D-(p+4))+1\;,\\
&a_{2}=-4(D-(p+1))(D-(p+4))\;,\\
&a_{3}=-2(D-(p+1))(D-(p+2))(D-(p+6))\;,\\
&a_{4}=-(D-(p+1))(D-(p+2))[D^{2}+16+p(p+9)-D(2p+9)]-1\;.
\end{split}
\eeq

Let's now expand the action (\ref{novelstv1}) to leading order in $(D-(p+1))\phi$, only keeping terms which will not vanish in the limit $D\to p+1$,
\beq
\begin{split}
I=&-\frac{1}{2\lambda^{p-1}_{s}}\int d^{p+1}x\sqrt{-g}e^{-\phi}\biggr\{R+a_{1}(\nabla\phi)^{2}\\
&-\frac{\alpha'}{4}\left[\mathcal{L}_{GB}+a_{2}G^{ab}(\nabla_{a}\phi)(\nabla_{b}\phi)+a_{3}(\nabla\phi)^{2}\Box\phi+a_{4}(\nabla\phi)^{4}\right]\\
&-\frac{\alpha'}{4}(D-(p+1))\phi\left[\mathcal{L}_{GB}+a_{2}G^{ab}(\nabla_{a}\phi)(\nabla_{b}\phi)+a_{3}(\nabla\phi)^{2}\Box\phi+a_{4}(\nabla\phi)^{4}\right]\biggr\}\;.
\end{split}
\label{anaction2}\eeq
If we were to now rescale the string coupling $\alpha'\to\frac{\alpha'}{(D-(p+1))}$, we find two divergences appear as $D\to p+1$, namely, 
\beq I_{div}=\frac{1}{2\lambda^{p-1}_{s}}\int d^{p+1}x\sqrt{-g}e^{-\phi}\frac{\alpha'}{4(D-(p+1))}[\mathcal{L}_{GB}-(\nabla\phi)^{4}]\;.\label{Idiv}\eeq
We recognize $I_{div}$ as precisely the $\alpha'$ correction to the bosonic string action in $p+1$ dimensions. Therefore, if we subtract $I_{div}$ from (\ref{anaction2}), rescale $\alpha'\to\frac{\alpha'}{(D-(p+1))}$ and take $D\to p+1$, we arrive to the `regularized' action:
\beq
\begin{split}
I_{reg}=&-\frac{1}{2\lambda^{p-1}_{s}}\int d^{p+1}x\sqrt{-g}e^{-\phi}\biggr\{R+(\nabla\phi)^{2}-\frac{\alpha'}{4}\phi\left[\mathcal{L}_{GB}-(\nabla\phi)^{4}\right]\\
&-\frac{\alpha'}{4}\left[12G^{ab}(\nabla_{a}\phi)(\nabla_{b}\phi)-10(\nabla\phi)^{2}\Box\phi+8(\nabla\phi)^{4}\right]\biggr\}\;.
\end{split}
\label{anaction2}\eeq

Unlike dimensionally reduced pure Einstein-Gauss-Bonnet gravity, where, at least in $p+1\leq4$ dimensions, a purely topological contribution (\ref{gbsub}) is subtracted, here $I_{div}$ (\ref{Idiv}) would generally alter the equations of motion of the $p+1\leq4$. This observation, coupled with a similar difficulty outlined in \cite{Easson:2020mpq}, suggests a novel limit of the string action in question is not natural and ill-posed.

Alternatively, at first glance it seems natural to Taylor expand the complete conformal factor $e^{(D-(p+2))\phi}$ to order $O(D-(p+2))$, and then subtract the topological contribution $\sim\int d^{p+1}x\sqrt{-g}\frac{\alpha'}{4}\mathcal{L}_{GB}$. The cost of this, however, is that additional divergences arise from other contributions to the action (\ref{novelstv1}) in the limit $D\to p+2$, requiring a less physically motivated subtraction scheme. We also point out that novel EGB theory as described in \cite{Lu:2020iav} has $p+1\leq3$- and $p+2\leq2$-dimensional limits, due to the form of the coefficients $a_{i}$. In contrast, the coefficients in (\ref{novelstv1}) make such limits untenable, lacking a well-motivated subtraction scheme.

It is noteworthy to compare the regularized novel string action (\ref{anaction2}) to the novel string action given in Eq. (56) of \cite{Easson:2020mpq}. In \cite{Easson:2020mpq}, the novel action was arrived at by performing a dimensional regularization of the string action (\ref{stringact1}), where one subtracts from (\ref{stringact1}) a conformally transformed $\alpha'$ correction in $D$-dimensions and then rescales the $\alpha'$ coupling as done above. The resulting action led to further divergences, however, which could only be resolved by rescaling the dilaton as $\phi\to(D-4)\phi$. Consequently, the resulting regularized novel string action did not have any kinetic terms for the dilaton at tree level ($\alpha'=0$). This is in contrast with the novel action described here (\ref{anaction2}), which maintains kinetic terms for the dilaton at tree level. 

Lastly, the comparison between (\ref{anaction2}) and the action in \cite{Easson:2020mpq} is really a comparison between the two methods of writing down a novel theory: dimensional reduction versus the Ross-Mann dimensional regularization \cite{Mann:1992ar}. In the case of pure EGB, these two methods, upon trivial field redefinitions, are equivalent. In the string action context explored here, the two mechanisms are obviously different. Indeed, while our novel string action (\ref{anaction2}) is not shift symmetric, again due to the overall $e^{\phi}$ factor, the dimensionally regularized novel string action of \cite{Easson:2020mpq} is shift symmetric. Both definitions of the novel string action, however, lack a physically well-motivated subtraction scheme, tentatively suggesting novel limits are only feasible for pure theories of Lovelock gravity.



\section{Discussion and Conclusion} \label{conclusion}
\indent 

Realizing that (generalized) Galileons naturally emerge from Lovelock actions \cite{VanAcoleyen11-1}, we have gone further and demonstrated this subclass of theories is in fact embedded in 10-dimensional string theory. More specifically, truncating the gravi-dilaton sector of $D$-dimensional superstring theory to the $\alpha'$ correction, we found that a consistent diagonal dimensional reduction of this theory generically leads to a Horndeski theory of gravity. The coefficients are fixed by the parameters $\alpha,\beta$ in the KK metric ansatz, and the dimension of the spacetime $D=p+1+n$. Notably, depending on the choices of coefficients, the Horndeski theory found from dimensional reduction yields certain well known scalar-tensor theories of gravity, including the self-tuning Fab four, K-essence, and Brans-Dicke. Then,  fixing $D=10$, and working in the Einstein frame, we demonstrated for $(p,n)=(3,6)$ and $(p,n)=(2,7)$ there are two allowed actions $I_{\pm}$ corresponding to the pairs $(\alpha_{+},\beta_{+})$ and $(\alpha_{-},\beta_{-})$, respectively, where $\alpha_{\pm},\beta_{\pm}$ are imposed from working in Einstein frame. Also for a specific choice of $\alpha$ and $\beta$, we were able to write down an $\alpha'$ correction to a CGHS-like 1+1 model of gravity. 

We also asked whether a novel limit of the $D$-dimensional $\alpha'$-corrected effective string action was possible. Following \cite{Lu:2020iav} this was accomplished by a combination of dimensional reduction and the rescaling $\alpha'\to\alpha'/(D-(p+1))$. To avoid potential divergences arising in the limit $D\to p+1$, we saw we must subtract off precisely the $\alpha'$ correction to the effective string action specific to $p+1$-dimensions. This is similar to what is observed for novel EGB \cite{Lu:2020iav}, where divergences are eliminated by subtracting a pure Gauss-Bonnet Lagrangian in $p+1$-dimensions. Unlike the novel EGB, however, the $p+1$-dimensional term we must subtract is not purely topological in $p\leq3$ dimensions. Thus, without a physical motivation for this subtraction, it appears the novel limit of the Horndeski model studied here is \emph{ad hoc} and ill-posed. Moreover, even when the subtraction is justified, the novel string action written here is markedly disparate from the one derived using the dimensional regularization techniques in \cite{Easson:2020mpq}, demonstrating the differences between the dimensional regularization and dimensional reduction schemes.

Let us now discuss some interesting implications of our work, and point to avenues worthy of additional study.

\vspace{2mm}

\noindent \textbf{Consistency of the Null Energy Condition}

The classical null energy condition (NEC),
\beq T^{matter}_{ab}k^{a}k^{b}\geq0\;,\label{nec}\eeq
where $k^{a}$ is any null vector, imposes a lower bound on the energy-momentum tensor of \emph{matter}. For matter described by a perfect fluid with energy density $\rho$ and pressure $p$, the NEC (\ref{nec}) translates to $\rho+p\geq0$. In general relativity the NEC is connected to the purely geometric Ricci convergence condition, $R_{ab}k^{a}k^{b}\geq0$, via Einstein's field equations. Correspondingly, in an FLRW background with Hubble parameter $H=\frac{\dot{a}}{a}$, the NEC (\ref{nec}) is equivalent to\footnote{We should emphasize here we are assuming a spatially \emph{flat} FRLW background. Indeed, in a background with spatial curvature $k$, one has $\dot{H}=-4\pi G(\epsilon+p)+\frac{k}{a^{2}}$, such that when the NEC holds, $\epsilon+p\geq0$, for $k=+1$ and sufficiently small $a$ one has $\dot{H}>0$. We thank an anonymous referee for pointing this out to us.}
\beq \dot{H}\leq0\;.\,\label{nechubble}\eeq
Ultimately, it is the Ricci convergence condition that has been used in a variety of important proofs in classical GR, \emph{e.g.}, the singularity theorems of Penrose and Hawking. The NEC (\ref{nec}) itself, it should be emphasized, is often a condition imposed on matter so as to preclude the existence of `exotic' spacetimes such as traversable wormholes, time machines, and warp drives. More generally, it was argued in \cite{Sawicki:2012pz}, for systems which exhibit Lorentz invariance and whose Hamiltonian coincides with the energy density given by the energy-momentum tensor, that a violation of the NEC implies the phase space contains configurations with arbitrarily negative energy.\footnote{While the argument in \cite{Sawicki:2012pz} is a compelling reason why the NEC should hold -- to preclude unbounded negative energies in classical systems -- the NEC lacks a precise origin in classical gravity. Recent work, however, has shown the NEC can be derived from the second law of (spacetime) thermodynamics \cite{Parikh:2015ret,Parikh:2016lys}, or from the Virasoro constraint of a closed bosonic string propagating on a curved background at tree level $(\alpha'=0)$ \cite{Parikh:2014mja}, such that a violation of the NEC corresponds to a violation in the second law or the Virasoro constraint, respectively.}

The NEC also plays a pertinent role in cosmology. Standard models of inflation described by an Einstein-Hilbert action minimally coupled to a canonical scalar field naturally satisfy the bound (\ref{nec}), however, are known to be past incomplete, thereby motivating alternatives to inflation, such as non-singular bouncing cosmologies (c.f. \cite{Brandenberger:2009jq}). Bouncing cosmologies are known to violate the NEC, at least for some time interval. Such solutions exist in generic scalar-tensor theories, whereby the NEC one means (\ref{nechubble}), as the matter energy-momentum tensor is typically replaced with a tensor that is a non-minmally coupled combination of matter and geometry, \emph{i.e.}, $T_{ab}\neq\frac{\delta I^{matter}}{\delta g^{ab}}$, such that (\ref{nec}) and (\ref{nechubble}) are no longer directly linked.  For such generic scalar-tensor models, moreover, the NEC-violating stages of the bouncing cosmology are often plagued with gravitational instabilties. Horndeski theory, however, will typically remain stable as it violates the Hubble condition (\ref{nechubble}) \cite{Qiu:2011cy,Easson:2011zy}. Thus, Horndeski gravity is attractive not only because it is a higher derivative theory without ghost-like instabilities, but also violates the NEC (\ref{nechubble}) in a natural and typically stable way.

Those who prefer the NEC to be a statement purely about the matter living on the spacetime, as in (\ref{nec}), may call into question whether the Galileons truly violate the NEC. A nice feature of deriving Horndeski gravity from dimensional reduction of pure Einstein-Gauss-Bonnet gravity \cite{VanAcoleyen11-1} is that violations of the NEC (\ref{nec}), in principle, can be traced back to a condition on the matter living in the higher dimensional parent theory. To illustrate this point, consider EGB theory in $D$-dimensions, with a minimally coupled scalar field $\Phi$ in Einstein frame\footnote{Already there is confusion between energy conditions in Einstein frame versus string frame. This issue was studied and remedied in \cite{Chatterjee:2012zh} by modifying the string frame NEC such that it maps to the usual NEC obeyed in Einstein frame. Incidentally, both the modified string frame NEC and Einstein frame NEC obey the second law of thermodynamics, eluding to the result discovered in \cite{Parikh:2015ret,Parikh:2016lys}.}
\beq I=\frac{1}{16\pi G_{D}}\int d^{D}x\sqrt{-\hat{g}}\left(\hat{R}-\frac{1}{2}(\hat{\nabla}\Phi)^{2}-V(\Phi)+\alpha_{GB}\hat{G}\right)\;.\eeq
It is straightforward to show for a null vector $k^{a}$, the NEC (\ref{nec}) is manifestly satisfied, $T^{mat}_{MN}=T^{\Phi}_{MN}k^{M}k^{N}=(k\cdot\hat{\nabla}\Phi)^{2}\geq0$. The condition (\ref{nechubble}) may continue to hold, but by way of imposing conditions on an effective energy momentum tensor $T^{eff}_{MN}$ that cleanly separates  matter and geometric contributions, such that the sign of $T^{mat}_{MN}k^{M}k^{N}$ will influence the sign of $\dot{H}$ in an unambiguous way. Upon dimensionally reducing and transforming the lower dimensional system into Einstein frame,  however, one finds a  Horndeski model where the effective energy-momentum tensor $T^{eff}_{ab}$ can no longer be cleanly separated into matter and geometric contributions. Nonetheless, the sign of $\dot{H}$, which depends on conditions imposed on $T^{eff}_{ab}$, can be traced back to conditions imposed on $T^{eff}_{MN}$, and $T^{mat}_{MN}$. Therefore, in a sense, there is a consistency in energy conditions between the higher dimensional parent theory and reduced theory. 

By contrast, our derivation of the Galileons from a low energy effective string action obscures the consistency in energy conditions. The reason for this is our starting point is an action already with non-minimal couplings between the dilaton and curvature, even in Einstein frame. Thus, while a consistency check between the higher-dimensional $T^{eff}_{MN}k^{M}k^{N}\geq0$ and its lower dimensional counterpart may exist, neither have energy conditions which directly relate to a condition purely on matter living on the background. It would be interesting to explore the consistency of energy conditions as outlined above in more detail, where perhaps one may find energy conditions should be replaced with some other principle. 

\vspace{2mm}

\noindent \textbf{Superluminality and Swampland criterion}

Generalized Galileons belong to a much larger arena of low energy effective field theories (EFTs). Finding consistency constraints is therefore crucial for determining which theories should be ruled out on fundamental and observational grounds. Notably, whether the EFT has a consistent UV completion\footnote{By UV completion we mean an EFT which arises as the IR limit of a well-defined renormalizable quantum field theory or string theory, assumed to obey standard axioms of a theory with an S-matrix, namely, unitarity, microcausality, and locality.} to a full model is of fundamental importance. Criteria imposed by string theory, or quantum gravity more generally, were hoped to be stringent enough to select a specific, unique EFT whose UV completion is the specific theory of quantum gravity.  Post the second string revolution, however, the stringy constraints became vacuous, leading to an obscenely vast landscape of unconstrained low energy EFTs, navely suggesting string theory allows for any low-energy EFT be completed in the UV. Much effort has gone into uncovering consistency conditions which constrain this `anything goes' philosophy. Two sets of consistency conditions -- positivity constraints and swampland conjectures -- have emerged as useful ways to reduce the number of EFTs. Let us briefly comment on these consistency conditions and how they relate to the Horndeski model we have uncovered here. 

The first set of consistency conditions are positivity constraints, \emph{i.e.}, the signs of certain higher-dimensional operators in a theory characterized by a local, Lorentz invariant Lagrangian must all be strictly positive \cite{Adams:2006sv,Shore:2007um}.\footnote{For example, consider the leading low-energy (irrelevant) interaction Lagrangian of a massless scalar field $\pi$ with constant shift symmetry, $\mathcal{L}=(\partial\pi)^{2}+\frac{c_{3}}{\Lambda^{4}}(\partial\pi)^{4}$, where $\Lambda$ is some mass scale. As argued in \cite{Adams:2006sv}, the coefficient $c_{3}$ must be strictly positive, $c_{3}>0$, if the UV completion of an EFT with such an interaction term respects the usual axioms of the S-matrix. This is indeed the case for a linear sigma model, where $\pi$ is a Goldstone boson joined together with a Higgs field to form a complex scalar field.} The signs of the leading higher derivative interactions, unlike the signs of kinetic terms, are not associated with instabilities of the low-energy vacua, but rather the speed of fluctuations around non-trivial generic backgrounds become superluminal, conflicting with principles of causality and locality. The authors of  \cite{Easson:2013bda,Dobre:2017pnt,deRham:2021fpu} argued that perturbations of the Galileon field will propagate superluminally (specifically when external matter is included), and, consequently, it is expected these theories cannot be UV-completed in the standard Wilsonian sense.\footnote{It's possible such theories undergo ``classicalization" \cite{Dvali:2010jz,Dvali:2010ns}, such that the EFT UV completes itself by introducing extended field configurations that play the role of quantum excitations that slowly decay into soft IR elementary excitations.} Likewise, our string induced model of Galileons, from the lower-dimensional perspective, is expected to suffer from the same problems of superluminality. On the other hand, we know our model comes from a consistent dimensional reduction of UV complete string theory.

This begs the question\footnote{We thank an anonymous referee for pointing out this question to us, and encouraging us to think about its potential resolution.}:  how is it that the subclass of generalized Galileons we uncovered here, which originate from a UV complete theory in the standard sense, also exhibit superluminal propagation? First, it is important to recognize the gravi-dilaton sector of higher $\alpha'$ corrections are unlikely to ameliorate the situation, as such higher derivative corrections are thought to contribute higher order Galileon contributions that are perturbatively smaller than the $\alpha'$ correction, thus resulting in smaller and ultimately inconsequential deviations from the leading superluminal behavior. Thus, we expect the issue of superluminal propagation will persist at all orders of $\alpha'$. Second, we emphasize we have not included external matter or the Kalb-Ramond contributions to our model. It is certainly possible including the Kalb-Ramond tensor fields, which appear at all levels in the $\alpha'$ expansion and couple to the dilaton in a non-trivial way, will lead to subluminal propagating modes. Indeed, as we have noted before, the low-energy effective action of the parent string theory is only consistent with worldsheet conformal invariance -- a crucial ingredient showing the UV-completeness of string theory-- when the Kalb-Ramond field is explicitly included. It would be interesting and worthwhile to study this question in detail.

The second set of consistency conditions are known as the swampland conjectures \cite{Vafa:2005ui,Ooguri:2006in} (see \cite{Palti:2019pca,vanBeest:2021lhn} for recent reviews). These conjectures include \emph{e.g.}, the distance conjecture, weak gravity conjecture, and no global symmetries, each of which are used to show that not all consistent looking EFTs can be coupled to gravity in a consistent manner and be UV complete. All EFTs which violate these criteria do not admit a (string theoretic) UV completion, and are said to live in the swampland (as opposed to the string landscape). Thus, a natural question is: is Horndeski gravity a part of the swampland, or string landscape? As laid forth in \cite{Heisenberg:2019qxz,Brahma:2019kch}, subclasses of generalized Galileons, in particular models of Quintessence, become highly constrained by the so-called de Sitter conjecture  \cite{Agrawal:2018own,Obied:2018sgi}, where the gradient of a scalar field potential is bounded from below, ruling out metastable de Sitter solutions. Importantly, while constrained, these subclasses of Horndeski models may still exist in the habitable landscape. Since our string induced model of Galileons emerges from a UV complete string theory, it would be interesting study how it avoids the swampland. Such an investigation may lead to additional swampland criteria which could be used to constrain larger classes of Horndeski theories. We leave this for future work.

\vspace{2mm}

\noindent \textbf{Multi-Galileons and Beyond Horndeski}

In this article we solely focused on deriving the generalized Galileons from a dimensional reduction of a low energy effective string action. Note our reduced action is not the most general Horndeski theory we could write down; indeed we are missing the triple derivative contributions which arise from dimensionally reducing higher order Lovelock actions. It is reasonable to think an effective string action with higher order Lovelock contrbutions will result in the triple derivative term upon dimensional reduction. Such a term, however, will not appear at the $\alpha'$ level as we are guaranteed any higher derivatives to the effective string action must appear beyond $\alpha'$ corrections \cite{Gross:1986mw}.\footnote{Going beyond first order $\alpha'$ corrections would be interesting. For recent work in this regard, see \cite{Wang:2020eln}.}

Nonetheless, there are generalizations to Horndeski gravity which might be accessible via a dimensional reduction of an $\alpha'$-corrected string action. For example, multi-field Galileons \cite{Padilla:2010de,Padilla:2010ir,Deffayet:2010zh,Hinterbichler:2010xn} might arise by keeping additional modes of the Kaluza-Klein tower in such a way that, upon reducing, we would have a scalar-tensor model with multiple interacting scalar fields with second order equations of motion. Alternatively, dimensional reducing $D=11$ supergravity, the low-energy limit of M-theory, over a torus leads to a lower dimensional theory with multiple interacting scalar fields, where each dilaton $\phi_{i}$ corresponds to each circle comprising the torus. Whether the resulting theory describes multi-Galileons, depends on whether any higher curvature corrections may be added to the parent theory. 

Moreover, while we did not include the Kalb-Ramond field $B$ in our analysis, keeping the $B$-field may lead to additional scalar fields in reduced action. More precisely, by allowing the $B$-field to have components along the directions of the compactified internal space, each leg may contribute an additional scalar. An example of this is seen in \cite{Bakhmatov:2019dow}, where reducing a 10-dimensional supergravity action, at tree level, over a 2-torus will lead to a reduced action with four scalar fields. It would be interesting to consider a similar action with an $\alpha'$ correction, perform a general KK reduction and then study the scalar contributions, where we would find multi-Galileon contributions without needing the additional modes in the KK tower\footnote{We thank Eoin Colgain for pointing out this interesting direction to us.}

Lastly, recall an important feature of Horndeski gravity is that, due to their second-order equations of motion, they are manifestly free of any Ostrogradsky ghost instabilities. In theories with multiple fields, even just a scalar field and metric, it turns out the second-order equations of motion condition is not necessary for the absence of ghost instabilities. Rather, it is possible to construct theories with two fields or more with higher than second order equations of motion, yet, via a degeneracy condition, leads to a constrained theory with only second order equations of motion (see \cite{Kobayashi:2019hrl} for a recent review). Such (single) scalar-tensor models are referred to as degenerate higher order scalar tensor (DHOST) theories beyond Horndeski gravity \cite{Zumalacarregui:2013pma,Bettoni:2013diz}. It may be interesting to similarly search for a string theoretic origin of beyond Horndeski or the related models of mimetic gravity \cite{Chamseddine:2013kea,Langlois:2018jdg}. A first step would be to see whether the higher dimensional ``Beyond Lovelock'' theories \cite{Crisostomi:2017ugk} dimensionally reduce to a specific class of DHOST theories.

\vspace{2mm}

\noindent \textbf{Cosmology, Black Holes and Beyond}

We noticed our string inspired Horndeski model has some features which set it apart from its pure EGB cousins. Specifically, due to the overall dilaton factor present in the higher dimensional string theory -- which cannot be conformally scaled away -- the Horndeski model developed here is not symmetric under constant shifts to the dilaton. We therefore expect hairy black holes and hairy stars to exist as solutions. Moreover, modulo the points made about energy conditions above, our Horndeski model is of the type which naturally violates $\dot{H}\leq0$, so as to offer an alternative to inflationary cosmology. Lastly, here we have not explicitly included a cosmological constant or a dilaton potential at tree level. Including such contributions would be straightforward and should lead to (a)dS vacua. In fact, since our model includes the Fab Four as a subclass, we expect a local Minkowski vacuum for any value of the net bulk cosmological constant. A preliminary investigation of such solutions is currently underway \cite{sveskomantoneassontba}.

\vspace{4mm}

\noindent \textbf{Top-Down Holography}

In this article we focused on a diagonal reduction of a low energy effective string action with an $\alpha'$ correction, resulting in a scalar-tensor theory of gravity. More generally one can perform a non-diagonal dimensional reduction resulting in a theory where a Maxwell gauge field couples non-trivially to both the metric and dilaton. Specifically, taking a $D+1$-dimensional metric ansatz to be
\beq d\hat{s}^{2}_{D+1}=e^{2\alpha\phi}g_{\mu\nu}dx^{\mu}dx^{\nu}+e^{2\beta\phi}(A_{\mu}dx^{\mu}+dz)\;,\eeq
curling up the extra $z$-direction into a circle leads to the following $D$-dimensional Ricci scalar
 \beq
 \begin{split}
 \hat{R}&=e^{-2\alpha\phi}R-e^{-2\alpha\phi}\biggr\{2[\beta+\alpha(D-1)]\Box\phi+\left[\alpha^{2}(D-1)(D-2)+2\beta(\beta+(D-2)\alpha)\right](\nabla\phi)^{2}\biggr\}\\
 &-\frac{1}{4}e^{2(\beta-2\alpha)\phi}F^{2}\;,
 \end{split}
 \label{redEinH}\eeq
while the Gauss-Bonnet Lagrangian reduces to\footnote{The non-diagonal reduction of the Gauss-Bonnet term was first carried out in \cite{Buchdahl:1979wi,MuellerHoissen:1989yv}, though with a different metric ansatz. An expression for the reduced Gauss-Bonnet term was written down in Eq. (4.3) of \cite{Charmousis12-1}, which, while we agree with the overall structure, our dimensionful coefficients $a_{i}$ don't appear to match.}
\beq
\begin{split}
\hat{\mathcal{L}}_{GB}&=e^{-4\alpha\phi}\biggr(\mathcal{L}_{GB}+a_{1}G^{ab}(\nabla_{a}\phi)(\nabla_{b}\phi)+a_{2}(\nabla\phi)^{4}+a_{3}(\nabla\phi)^{2}\Box\phi\biggr)\\
&+e^{2(\beta-3\alpha)\phi}\biggr\{-\frac{1}{2}P_{abcd}F^{ab}F^{cd}+a_{4}F^{ab}[(\nabla^{c}\phi)\nabla_{c}F_{ab}-2(\nabla_{b}\phi)\nabla^{c}F_{ab}]+a_{5}F^{2}(\nabla\phi)^{2}\\
&+a_{6}(F_{bc}\nabla^{b}\phi)^{2}\biggr\}+\frac{3}{16}e^{4(\beta-2\alpha)\phi}\left(F^{4}+2F^{ad}F^{bc}F_{ab}F_{cd}\right)\;,
\end{split}
\label{reduxGBredcirc3}\eeq
 \beq P_{abcd}=R_{abcd}+g_{ad}R_{bc}-g_{ac}R_{bd}-g_{bd}R_{ac}+g_{bc}R_{ad}+\frac{1}{2}R(g_{ac}g_{bd}-g_{ad}g_{bc})\label{doubledualrie}\;.\eeq
Here the $a_{i}$ are coefficients which depend on the spacetime dimension $D$ and parameters $\alpha$ and $\beta$. Thus, a pure theory of Einstein-Gauss-Bonnet gravity has become a complicated Einstein-Maxwell-Dilaton (EMD) theory of gravity. The scalar-vector-tensor model can be understood as a generalized Horndeski model including vector-tensor and vector-scalar interactions.\footnote{While known primarily for higher derivative theories of scalar-tensor gravity, Horndeski actually worked out the most general minimal coupling of a Maxwell field yielding second order equations of motion with higher derivatives \cite{Horndeski:1976gi}. Unsurprisingly, such a model arises from a Kaluza-Klein reduction of a Gauss-Bonnet term where the metric ansatz includes the $U(1)$ vector field $A_{\mu}$ interpreted as a Maxwell gauge field \cite{Buchdahl:1979wi}.}

Dimensionally reduced Einstein-Gauss-Bonnet gravity in the form of (\ref{redEinH}) and (\ref{reduxGBredcirc3}) is of particular interest to the study of holographic condensed matter systems,  specifically the holographic dictionary of EMD theories (c.f. \cite{Taylor:2008tg,Charmousis:2010zz,Lee:2010ii,Liu:2010ka}). As shown in \cite{Gouteraux:2011qh}, a particular class of EMD theories can be `oxidized' to a higher dimensional AdS-Maxwell gravity theory, which upon dimensional reduction leads to the desired EMD theory. Consequently, the complicated charged dilatonic black hole solutions of the lower dimensional EMD theory can be directly related to the simpler higher dimensional black hole and black brane solutions of AdS-Maxwell gravity. An Einstein-Gauss-Bonnet action of the form (\ref{redEinH}) and (\ref{reduxGBredcirc3}) was previously analyzed in \cite{Charmousis12-1} for a specific choice of metric ansatz parameters $\alpha$ and $\beta$, for which the hydrodynamic and thermodynamic  properties of the black hole solutions were extensively studied. It may be interesting to consider different choices of $\alpha$ and $\beta$.

A non-diagonal reduction of a low energy effective string action with $\alpha'$ corrections may also be of interest to the AdS/CMT community. Broadly, there are two methods in constructing models of holographic quantum matter: `bottom-up' and `top-down'. In the former often one starts with an EMD action that realizes solutions with Lifshitz scaling and hyperscaling violation and are `bottom-up' in that it is unclear whether such EMD actions emerge from a higher dimensional string/brane theory with well-defined UV completion. Alternatively, a `top-down' construction would be to start with a UV complete string theory and then study the resulting solutions of the theory. For example, a non-diagonal KK reduction of pure Einstein gravity in AdS leads to actions with hyperscaling violating solutions\footnote{Higher derivative theories of gravity and theories with additional matter fields also admit such solutions, see, e.g., \cite{Gath:2012pg,OKeeffe:2013xdv,Li:2016rcv,Pedraza:2018eey}.} \cite{Charmousis:2010zz,Gouteraux:2011ce}. Embedding the $\alpha'$ effective string action considered here and performing a non-diagonal reduction would lead to a type of generalized EMD theory, in a `top-down' sense, which may admit new hyperscaling violating solutions. It would be interesting to explore this further.

\vspace{2mm}

\noindent \textbf{$p$-form Galileons}

Finally, in this article we considered the gravi-dilaton sector of the $\alpha'$-corrected effective string action. Of course, maintaining conformal invariance of the worldsheet string action demands we include the 3-form gauge curvature $H=dB$ of the antisymmetric Kalb-Ramond field $B_{\mu\nu}$.
Thus, generally, for complete consistency, we should really be starting with the full effective string action including all of its sectors, of which we then dimensionally reduce. It is simple enough to reduce the additional $p$-form fields. For example, given a $p$-form field strength $\hat{F}_{(p)}$ contributing to the $D+1$-dimensional bosonic sector as $\mathcal{L}_{gauge}=-\frac{\sqrt{-\hat{g}}}{2p!}\hat{F}^{2}_{(p)}$, KK reduction leads to the sum of two dilaton-$p$-form interactions in $D$-dimensions \cite{Duff86-1}
\beq \mathcal{L}_{gauge}\to\sqrt{-g}\left(\frac{1}{2p!}e^{-2(p-1)\alpha\phi}F^{2}_{(p)}-\frac{1}{2(p-1)!}e^{2(D-p)\alpha\phi}F^{2}_{(p-1)}\right)\;.\eeq

Naturally then, an interesting future direction would be to derive the general $p$-form Galileons \cite{Deffayet:2010zh} arising via a Kaluza-Klein reduction of the $\alpha'$ corrected superstring action by including its $p$-form gauge field sectors. Provided the higher $p$-form gauge fields in the $\alpha'$ correction can be made to come in such a way that their equations of motion are second order (see, \emph{e.g.}, Eq. (2.34) of \cite{Metsaev:1987zx}), a non-diagonal reduction will result in a complicated theory of $p$-form Galileons. The expectation is the various sectors in the string theory will produce specific $p$-form Galileons. As we uncovered the (scalar) Galileons via diagonally reducing the gravi-dilaton sector of the superstring action -- with additonal vector Galileons from a non-diagonal reduction -- reducing  the Kalb-Ramond-dilaton sector will result in $3$-form Galileons interacting with the dilaton $\phi$, vector $A_{\mu}$, and the metric. In other words, the $p$-form Galileons in the reduced theory will be neatly organized based on which sector of the parent string theory they originate from. 

Moreover, thus far we have only concerned ourselves with the bosonic sector of the low energy effective superstring action. Of course, as a superstring theory, the action includes a fermionic sector comprised of the superpartners to the bosonic sector, \emph{e.g.}, the gravitino. Provided one has the fermionic analog of the $\alpha'$ correction to the bosonic sector including the Gauss-Bonnet term,  dimensionally reducing the $\alpha'$ correction to the fermionic sector is expected to result in supersymmetric Galileons. Such a computation would not only generalize the supersymmetrization
 presented in \cite{Elvang:2017mdq}, but also provide a more fundamental explanation for supersymmetric Galileons. 


\noindent\section*{ACKNOWLEDGMENTS}

We are pleased to thank Saugata Chatterjee, Blaise Gout\'eraux, Hong Lu, Juan Pedraza, and George Zahariade for helpful and illuminating discussions. We also thank an anonymous referee for their detailed comments which have greatly improved the outlook of this work. DE is supported in part by a grant from FQXi. AS is funded by the Simons Foundation \emph{It from Qubit} collaboration (Oppenheim).

\appendix

\section{Kaluza-Klein Reduction of Einstein-Gauss-Bonnet} \label{appBKKhigher} 

For completeness and pedagogy, we present the Kaluza-Klein reduction of the Einstein-Hilbert and Gauss-Bonnet terms in $p+1+n$ dimensions to a $p+1$-dimensional spacetime. We dimensionally reduce over an $n$-dimensional internal space, whose isometry group is Abelian, such that the massive modes of the KK reduction can be consistently truncated. Our set-up follows \cite{Charmousis12-1,Charmousis:2014mia}, however, we will need the total derivatives they drop, which is another motivation for us to carry out the calculation explicitly. 

We take our metric ansatz for the $D=p+1+n$-dimensional spacetime to be:
\beq d\hat{s}^{2}_{p+1+n}=\hat{g}_{MN}dx^{M}dx^{N}=e^{2\alpha\phi}g_{\mu\nu}dx^{\mu}dx^{\nu}+e^{2\beta\phi}\tilde{g}_{ij}(dz^{i}+A^{(i)}_{\mu}dx^{\mu})(dz^{j}+A^{(j)}_{\nu}dx^{\nu})\;,\eeq
where $\{z^{i}\}$ denotes the coordinates of the internal space, with metric $\tilde{g}_{ij}$. We will use the vielbein formalism to compute all relevant curvature quantities. The index notation is as follows. Here we use $M,N,...$ to denote the world indices of the full $p+1+n$-dimensional spacetime; $A,B,...$ to denote the local (Lorentz) indices of $p+1+n$-dimensional spacetime; $\mu,\nu,...$ to denote the world indices of the $p+1$-dimensional spacetime; $a,b,...,$ the Lorentz indices of the $p+1$-dimensional spacetime, and $\tilde{a},\tilde{b},...$ and $i,j,..$ denote the Lorentz and world indices of the $n$-dimensional internal space, respectively.

Using $\hat{g}_{MN}=\hat{e}^{A}_{M}\hat{e}^{B}_{N}\eta_{AB}$, it is straightforward to workout 
\beq \hat{e}^{a}=e^{\alpha\phi}e^{a}\;,\quad \hat{e}^{\tilde{a}}=e^{\beta\phi}(e^{\tilde{a}}+A^{(\tilde{b})})\;,\quad A^{(\tilde{b})}=A^{(\tilde{b})}_{\mu}dx^{\mu}=A^{(j)}_{\mu}e^{\tilde{b}}_{j}dx^{\mu}\;. \eeq
 In order to find the $p+1$-dimensional curvature in terms of the higher-dimensional curvature, we compute the $p+1+n$-dimensional curvature 2-forms, $\hat{\Omega}^{A}_{\;B}$, from which we may find the Riemann curvature via
\beq \hat{\Omega}^{A}_{\;B}=\frac{1}{2}\hat{R}^{A}_{\;BCD}\hat{e}^{C}\wedge\hat{e}^{D}\;.\eeq
The curvature 2-forms are found from Cartan's second structure equation,
\beq \hat{\Omega}^{A}_{\;B}=d\hat{\omega}^{A}_{\;B}+\hat{\omega}^{A}_{\;C}\wedge\hat{\omega}^{C}_{\;B}\;,\label{cartans2ndeqn}\eeq
where we assume an antisymmetric, torsionless spin-connection $\hat{\omega}^{A}_{\;B}$. We must therefore first compute this spin-connection, which is determined by Cartan's first structure equation,
\beq d\hat{e}^{A}+\hat{\omega}^{A}_{\;B}\wedge\hat{e}^{B}=0\;.\eeq

We find
\beq
\begin{split}
&\hat{\omega}^{\tilde{a}}_{\;a}=\beta(\partial_{a}\phi)e^{-\alpha\phi}\hat{e}^{\tilde{a}}+\frac{1}{2}e^{(\beta-2\alpha)\phi}F^{(\tilde{a})}_{ab}\hat{e}^{b}-e^{(\beta-\alpha)\phi}\omega^{\tilde{a}}_{\;\tilde{b}}A^{(\tilde{b})}_{a}\;,\\
&\hat{\omega}^{a}_{\;\tilde{a}}=-\eta_{\tilde{a}\tilde{b}}\left[\beta(\partial^{a}\phi)e^{-\alpha\phi}\hat{e}^{\tilde{b}}+\frac{1}{2}e^{(\beta-2\alpha)\phi}F^{(\tilde{b})a}_{\;\;\;c}\hat{e}^{c}-e^{(\beta-\alpha)\phi}\omega^{\tilde{b}}_{\;\tilde{c}}A^{(\tilde{c})a}\right]\;,\\
&\hat{\omega}^{a}_{\;b}=\omega^{a}_{\;b}+\alpha e^{-\alpha\phi}(\partial_{b}\phi\hat{e}^{a}-\partial^{a}\phi\eta_{bc}\hat{e}^{c})-\frac{1}{2}e^{(\beta-2\alpha)}F^{a}_{\;(\tilde{a})b}\hat{e}^{\tilde{a}}\;,\\
&\hat{\omega}^{\tilde{a}}_{\;\tilde{b}}=\omega^{\tilde{a}}_{\;\tilde{b}}\;.
\end{split}
\eeq
Here $\omega^{a}_{\;b}$ is the spin-1 connection of the $p+1$-dimensional space while $\omega^{\tilde{a}}_{\;\tilde{b}}$ is the spin-1 form of the internal $n$-dimensional space. 

We will only be interested in the gravi-dilaton sector, such that from here on we drop all terms which include a gauge potential. Moreover, for ease we work in the limit the internal Euclidean space is flat, $\omega^{\tilde{a}}_{\;\tilde{b}}=0$. Then, we have
\beq \hat{e}^{a}=e^{\alpha\phi}e^{a}\;,\quad \hat{e}^{\tilde{a}}=e^{\beta\phi}e^{\tilde{a}}\;,\eeq
\beq  d\hat{e}^{a}=\alpha(\partial_{b}\phi)e^{-\alpha\phi}\hat{e}^{b}\wedge\hat{e}^{a}-\omega^{a}_{\;b}\wedge\hat{e}^{b}\;,\quad  d\hat{e}^{\tilde{a}}=\beta(\partial_{b}\phi)e^{-\alpha\phi}\hat{e}^{b}\wedge\hat{e}^{\tilde{a}}\;,\eeq
\beq
\begin{split}
& \hat{\omega}^{\tilde{a}}_{\;a}=\beta(\partial_{a}\phi)e^{-\alpha\phi}\hat{e}^{\tilde{a}}\;,\quad \hat{\omega}^{a}_{\;\tilde{a}}=-\eta_{\tilde{a}\tilde{b}}\beta(\partial^{a}\phi)e^{-\alpha\phi}\hat{e}^{\tilde{b}}\;,\\
&\hat{\omega}^{a}_{\;b}=\omega^{a}_{\;b}+\alpha e^{-\alpha\phi}\left[(\partial_{b}\phi)\hat{e}^{a}-(\partial^{a}\phi)\eta_{bc}\hat{e}^{c}\right]\;.
\end{split}
\eeq
Using Cartan's second structure equation, we work out the curvature 2-forms. The only non-zero terms are
\beq
\begin{split}
&\hat{\Omega}^{a}_{\;b}=\Omega^{a}_{\;b}+\frac{\alpha}{2} e^{-2\alpha\phi}\biggr\{[(\nabla_{c}\partial_{b}\phi)\delta^{a}_{\;d}-(\nabla_{d}\partial_{b}\phi)\delta^{a}_{c}]-[(\nabla_{c}\partial^{a}\phi)\eta_{bd}-(\nabla_{d}\partial^{a}\phi)\eta_{bc}]\\
&+\alpha\left((\partial_{b}\phi)[(\partial_{d}\phi)\delta^{a}_{\;c}-(\partial_{c}\phi)\delta^{a}_{\;d}]-(\partial\phi)^{2}(\eta_{bd}\delta^{a}_{\;c}-\eta_{bc}\delta^{a}_{\;d})+(\partial^{a}\phi)[(\partial_{c}\phi)\eta_{bd}-(\partial_{d}\phi)\eta_{bc}]\right)\biggr\}\hat{e}^{c}\wedge\hat{e}^{d}\;,
\end{split}
\eeq
and
\beq \hat{\Omega}^{a}_{\;\tilde{a}}=-\eta_{\tilde{a}\tilde{b}}\beta e^{-2\alpha\phi}\left[(\beta-2\alpha)(\partial^{a}\phi)(\partial_{c}\phi)+\alpha(\partial\phi)^{2}\delta^{a}_{\;c}+\nabla_{c}(\partial^{a}\phi)\right]\hat{e}^{c}\wedge\hat{e}^{\tilde{b}}\;.\eeq

Using $\hat{\Omega}^{A}_{\;B}=\frac{1}{2}\hat{R}^{A}_{\;BCD}\hat{e}^{C}\wedge\hat{e}^{D}$, the non-vanishing components of the Riemann tensor are:
\beq
\begin{split}
 \hat{R}^{a}_{\;bcd}&=e^{-2\alpha\phi}R^{a}_{\;bcd}+\alpha e^{-2\alpha\phi}\biggr\{[(\nabla_{c}\partial_{b}\phi)\delta^{a}_{\;d}-(\nabla_{d}\partial_{b}\phi)\delta^{a}_{\;c}]-[(\nabla_{c}\partial^{a}\phi)\eta_{bd}-(\nabla_{d}\partial^{a}\phi)\eta_{bc}]\\
&+\alpha\left((\partial_{b}\phi)[(\partial_{d}\phi)\delta^{a}_{\;c}-(\partial_{c}\phi)\delta^{a}_{\;d}]-(\partial\phi)^{2}(\eta_{bd}\delta^{a}_{\;c}-\eta_{bc}\eta^{a}_{\;d})+(\partial^{a}\phi)[(\partial_{c}\phi)\eta_{bd}-(\partial_{d}\phi)\eta_{bc}]\right)\biggr\}\;,
\end{split}
\eeq
\beq \hat{R}^{a}_{\;\tilde{a}c\tilde{b}}=-\eta_{\tilde{a}\tilde{b}}\beta e^{-2\alpha\phi}\left[(\beta-2\alpha)(\partial^{a}\phi)(\partial_{c}\phi)+\alpha(\partial\phi)^{2}\delta^{a}_{\;c}+\nabla_{c}(\partial^{a}\phi)\right]\;.\eeq

With the Riemann curvature in hand, the Ricci curvature components are
\beq
\begin{split}
\hat{R}_{bd}&=e^{-2\alpha\phi}R_{bd}-e^{-2\alpha\phi}\biggr\{(\alpha(p-1)+n\beta)[(\nabla_{d}\nabla_{b}\phi)+\alpha\eta_{bd}(\nabla\phi)^{2}]+\alpha\eta_{bd}\Box\phi\\
&+[n\beta(\beta-2\alpha)-\alpha^{2}(p-1)](\nabla_{b}\phi)(\nabla_{d}\phi)\biggr\}\;,
\end{split}
\eeq
\beq \hat{R}_{\tilde{b}\tilde{d}}=-\eta_{\tilde{b}\tilde{d}}\beta e^{-2\alpha\phi}\left[(\beta n+(p-1)\alpha)(\nabla\phi)^{2}+\Box\phi\right]\;.\eeq

We now have all of the ingredients necessary to work out the Ricci scalar
\beq \hat{R}=\eta^{AB}\hat{R}_{AB}=\eta^{bd}\hat{R}_{bd}+\eta^{\tilde{b}\tilde{d}}\hat{R}_{\tilde{b}\tilde{d}}\;.\eeq
Only a little algebra yields
\beq \hat{R}=e^{-2\alpha\phi}\biggr\{R-2(\alpha p+n\beta)\Box\phi-[\alpha^{2}p(p-1)+n(n+1)\beta^{2}+2n(p-1)\alpha\beta](\nabla\phi)^{2}\biggr\}\;.\eeq
Then, with $\sqrt{-\hat{g}}=e^{(n\beta+(p+1)\alpha)\phi}\sqrt{-g}$, we have
\beq \sqrt{-\hat{g}}\hat{R}=\sqrt{-g}e^{(\alpha(p-1)+\beta n)\phi}\biggr\{R-2(\alpha p+n\beta)\Box\phi-[\alpha^{2}p(p-1)+n(n+1)\beta^{2}+2n(p-1)\alpha\beta](\nabla\phi)^{2}\biggr\}\;.\label{ricscal}\eeq
Our expression for the Ricci scalar matches Eq. (A.3) of \cite{Charmousis12-1} in the flat space limit. 

\vspace{1mm}

\noindent \textbf{Gauss-Bonnet Curvature}

Moving on, let us now work out the Gauss-Bonnet curvature $\hat{\mathcal{L}}_{GB}=\hat{R}^{2}-4\hat{R}^{2}_{AB}+\hat{R}^{2}_{ABCD}$. The Ricci scalar is the easiest to work out:
\beq \hat{R}^{2}=e^{-4\alpha\phi}(R^{2}-2A_{1}\Box\phi R-2A_{2}(\nabla\phi)^{2}R+A_{1}^{2}(\Box\phi)^{2}+A_{2}^{2}(\nabla\phi)^{4}+2A_{1}A_{2}\Box\phi(\nabla\phi)^{2})\;,\eeq
with
\beq A_{1}\equiv2(\alpha p+n\beta)\;,\quad A_{2}\equiv\alpha^{2}p(p-1)+n(n+1)\beta^{2}+2n(p-1)\alpha\beta\;.\eeq

The Ricci tensor squared is composed of $\hat{R}^{2}_{AB}=\hat{R}_{ab}\hat{R}^{ab}+\hat{R}_{\tilde{a}\tilde{b}}\hat{R}^{\tilde{a}\tilde{b}}$. It is easy to work out
\beq \hat{R}_{\tilde{b}\tilde{d}}\hat{R}^{\tilde{b}\tilde{d}}=n\beta^{2}e^{-4\alpha\phi}\left[(\beta n+\alpha(p-1))^{2}(\nabla\phi)^{4}+(\Box\phi)^{2}+2(\beta n+\alpha(p-1))\Box\phi(\nabla\phi)^{2}\right]\;,\eeq
\beq 
\begin{split}
\hat{R}^{bd}\hat{R}_{bd}&=e^{-4\alpha\phi}\biggr\{R^{2}_{bd}-2A_{3}R^{bd}(\nabla_{b}\nabla_{d}\phi)-2\alpha A_{3}R(\nabla\phi)^{2}-2A_{4}R^{bd}(\nabla_{b}\phi)(\nabla_{d}\phi)-2\alpha R\Box\phi\\
&+A_{3}^{2}(\nabla\nabla\phi)^{2}+2A_{3}A_{4}(\nabla\nabla\phi)(\nabla\phi)(\nabla\phi)+[2A_{3}\alpha+\alpha^{2}(p+1)](\Box\phi)^{2}\\
&+[A_{3}\alpha(A_{3}\alpha(p+1)+2A_{4})+A_{4}^{2}](\nabla\phi)^{4}+[2A_{3}\alpha(A_{3}+\alpha(p+1))+2\alpha A_{4}](\Box\phi)(\nabla\phi)^{2}\biggr\}\;,
\end{split}
\eeq
where
\beq A_{3}\equiv(\alpha(p-1)+n\beta)\;,\quad A_{4}\equiv[n\beta(\beta-2\alpha)-\alpha^{2}(p-1)]\;.\eeq
Therefore, 
\beq 
\begin{split}
\hat{R}^{2}_{BD}&=e^{-4\alpha\phi}\biggr\{R^{2}_{bd}-2A_{3}R^{bd}(\nabla_{b}\nabla_{d}\phi)-2\alpha A_{3}R(\nabla\phi)^{2}-2A_{4}R^{bd}(\nabla_{b}\phi)(\nabla_{d}\phi)-2\alpha R\Box\phi\\
&+A_{3}^{2}(\nabla\nabla\phi)^{2}+2A_{3}A_{4}(\nabla\nabla\phi)(\nabla\phi)(\nabla\phi)+A_{5}(\Box\phi)^{2}+A_{6}(\nabla\phi)^{4}+A_{7}\Box\phi(\nabla\phi)^{2}\biggr\}\;,
\end{split}
\eeq
with
\beq 
\begin{split}
&A_{5}\equiv[2A_{3}\alpha+\alpha^{2}(p+1)+n\beta^{2}]\;,\\
&A_{6}\equiv[A_{3}\alpha(A_{3}\alpha(p+1)+2A_{4})+A_{4}^{2}+n\beta^{2}(\beta n+\alpha(p-1))^{2}]\;,\\
&A_{7}\equiv[2A_{3}\alpha(A_{3}+\alpha(p+1))+2\alpha A_{4}+2n\beta^{2}(\beta n+\alpha(p-1))]\;.
\end{split}
\eeq
 
Moving on, the Riemann curvature squared satisfies $\hat{R}^{2}_{ABCD}=\hat{R}_{abcd}^{2}+4\hat{R}_{a\tilde{b}c\tilde{d}}^{2}$, where
\beq 
\begin{split}
\hat{R}_{a\tilde{b}c\tilde{d}}^{2}&=n\beta^{2}e^{-4\alpha\phi}\biggr\{[(\beta-2\alpha)^{2}+2\alpha(\beta-2\alpha)+\alpha^{2}(p+1)](\nabla\phi)^{4}+2\alpha(\nabla\phi)^{2}\Box\phi\\
&+2(\beta-2\alpha)(\nabla\phi)(\nabla\phi)(\nabla\nabla\phi)+(\nabla\nabla\phi)^{2}\biggr\}\;,
\end{split}
\eeq
\beq 
\begin{split}
\hat{R}_{abcd}^{2}&=e^{-4\alpha\phi}\biggr\{R^{2}_{abcd}-8\alpha R^{ab}(\nabla_{a}\nabla_{b}\phi)+8\alpha^{2}G^{ab}(\nabla_{a}\phi)(\nabla_{b}\phi)\\
&+4\alpha^{2}(p-1)(\nabla\nabla\phi)^{2}+4\alpha^{2}(\Box\phi)^{2}-8\alpha^{3}(p-1)(\nabla\nabla\phi)(\nabla\phi)(\nabla\phi)+8\alpha^{3}(p-1)\Box\phi(\nabla\phi)^{2}\\
&+2\alpha^{4}p(p-1)(\nabla\phi)^{4}\biggr\}\;,
\end{split}
\eeq
such that
\beq
\begin{split}
\hat{R}^{2}_{ABCD}&=e^{-4\alpha\phi}\biggr\{R^{2}_{abcd}-8\alpha R^{ab}(\nabla_{a}\nabla_{b}\phi)+8\alpha^{2}G^{ab}(\nabla_{a}\phi)(\nabla_{b}\phi)\\
&+A_{8}(\nabla\nabla\phi)^{2}+A_{9}\Box\phi(\nabla\phi)^{2}+4\alpha^{2}(\Box\phi)^{2}+A_{10}(\nabla\nabla\phi)(\nabla\phi)(\nabla\phi)+A_{11}(\nabla\phi)^{4}\biggr\}\;,
\end{split}
\eeq
with
\beq
\begin{split}
&A_{8}\equiv4[\alpha^{2}(p-1)+n\beta^{2}]\;,\\
&A_{9}\equiv 8\alpha[\alpha^{2}(p-1)+n\beta^{2}]\;,\\
&A_{10}\equiv 8[n\beta^{2}(\beta-2\alpha)-\alpha^{3}(p-1)]\;,\\
&A_{11}\equiv[2\alpha^{4}p(p-1)+4n\beta^{2}[(\beta-2\alpha)^{2}+2\alpha(\beta-2\alpha)+\alpha^{2}(p+1)]]\;.
\end{split}
\eeq

Putting everything together, we have $\hat{\mathcal{L}}_{GB}$:
\beq 
\begin{split}
\hat{\mathcal{L}}_{GB}&=e^{-4\alpha\phi}\biggr\{\mathcal{L}_{GB}+C_{1}R^{ab}(\nabla_{b}\nabla_{b}\phi)+C_{2}R^{ab}(\nabla_{a}\phi)(\nabla_{b}\phi)+C_{3}G^{ab}(\nabla_{a}\phi)(\nabla_{b}\phi)+C_{4}R\Box\phi\\
&+C_{5}R(\nabla\phi)^{2}+C_{6}(\nabla\nabla\phi)(\nabla\phi)(\nabla\phi)+C_{7}(\nabla\nabla\phi)^{2}+C_{8}(\Box\phi)^{2}+C_{9}(\nabla\phi)^{4}+C_{10}\Box\phi(\nabla\phi)^{2}\biggr\}\;,
\end{split}
\label{Gaussbonnredint}\eeq
where
\beq 
\begin{split}
&C_{1}\equiv8(A_{3}-\alpha)\;,\quad C_{2}\equiv8A_{4}\;,\quad C_{3}\equiv8\alpha^{2}\;,\\
&C_{4}\equiv(8\alpha-2A_{1})\;,\quad C_{5}\equiv(8\alpha A_{3}-2A_{2})\;,\\
&C_{6}\equiv[A_{10}-8A_{3}A_{4}]\;,\quad C_{7}\equiv[A_{8}-4A_{3}^{2}]\;,\quad C_{8}\equiv[A_{1}^{2}+4\alpha^{2}-4A_{5}]\;,\\
&C_{9}\equiv[A_{2}^{2}+A_{11}-4A_{6}]\;,\quad C_{10}\equiv[2A_{1}A_{2}+A_{9}-4A_{7}]\;.
\end{split}
\eeq

Specifically, the constants $C_{i}$ are:
\beq C_{1}=8[n\beta+\alpha(p-2)]\;,\quad C_{2}=8n\beta(\beta-2\alpha)-8(p-1)\alpha^{2}\;,\quad C_{3}=8\alpha^{2}\;,\eeq
\beq C_{4}=-4[n\beta+\alpha(p-2)]=-\frac{1}{2}C_{1}\;,\eeq
\beq C_{5}=-2(p-1)(p-4)\alpha^{2}-4n(p-3)\alpha\beta-2n(n+1)\beta^{2}\;,\eeq
\beq C_{6}=8(p-2)(p-1)\alpha^{3}+24n(p-1)\alpha^{2}\beta-8n(p+1-2n)\alpha\beta^{2}-8n(n-1)\beta^{3}\;,\eeq
\beq C_{7}=-4\left[(p-1)(p-2)\alpha^{2}+2n(p-1)\alpha\beta+n(n-1)\beta^{2}\right]\;,\eeq
\beq C_{8}=4(p-2)(p-1)\alpha^{2}+8n(p-1)\alpha\beta+4n(n-1)\beta^{2}=-C_{7}\;,\eeq
\beq 
\begin{split}
C_{9}&=p(p-1)(p-2)(p-3)\alpha^{4}+4np(p-1)(p-3)\alpha^{3}\beta+2np(5-p+n(3p-7))\alpha^{2}\beta^{2}\\
&+4n(n-1)[2+n(p-1)]\alpha\beta^{3}+n(n-1)[-4+n(n-1)]\beta^{4}\;,
\end{split}
\eeq
\beq
\begin{split}
C_{10}&=4\left[(p-2)^{2}(p-1)\alpha^{3}+3n(p^{2}-3p+2)\alpha^{2}\beta+n(2-p+n(3p-4))+n^{2}(n-1)\beta^{3}\right]\;.
\end{split}
\eeq

As written, $\hat{\mathcal{L}}_{GB}$ does not manifestly return second order equations of motion due to the presence of terms such as $(\Box\phi)^{2}$. However, we can make additional simplifications such that we find second order equations of motion upon dropping total derivatives. In particular, using the identities
\beq (\Box\phi)^{2}-(\nabla\nabla\phi)^{2}=R^{ab}(\nabla_{a}\phi)(\nabla_{b}\phi)+\nabla_{a}[\Box\phi(\nabla^{a}\phi)-(\nabla^{a}\nabla^{b}\phi)(\nabla_{b}\phi)]\;,\label{useid1}\eeq
\beq (\nabla\nabla\phi)(\nabla\phi)(\nabla\phi)=-\frac{1}{2}\Box\phi(\nabla\phi)^{2}+\frac{1}{2}\nabla_{a}[(\nabla\phi)^{2}\nabla^{a}\phi]\;,\label{useid2}\eeq
and simplifying, we find
\beq \begin{split}
\hat{\mathcal{L}}_{GB}&=e^{-4\alpha\phi}\biggr\{\mathcal{L}_{GB}+(C_{3}-2C_{5})G^{ab}(\nabla_{a}\phi)(\nabla_{b}\phi)+C_{9}(\nabla\phi)^{4}+\left(C_{10}-\frac{C_{6}}{2}\right)\Box\phi(\nabla\phi)^{2}\biggr\}\;,
\end{split}
\label{Ghatscaltenhigh}\eeq
along with the total derivative term
\beq e^{-4\alpha\phi}\nabla_{a}\biggr\{C_{1}G^{ab}\nabla_{b}\phi+\frac{C_{6}}{2}(\nabla\phi)^{2}(\nabla^{a}\phi)-C_{7}[\Box\phi(\nabla^{a}\phi)-(\nabla^{a}\nabla^{b}\phi)(\nabla_{b}\phi)]\biggr\}\;.\label{totderivs}\eeq
Here
\beq C_{3}-2C_{5}=4[(p-2)(p-3)\alpha^{2}+2n(p-3)\alpha\beta+n(n+1)\beta^{2}]\;,\eeq
\beq 
\begin{split}
C_{10}-\frac{1}{2}C_{6}&=4\biggr\{(p-1)(p-2)(p-3)\alpha^{3}+3n(p-1)(p-3)\alpha^{2}\beta+3n(n(p-2)+1)\alpha\beta^{2}\\
&+n(n-1)(n+1)\beta^{3}\biggr\}\;.
\end{split}
\eeq

We care about the form of the total derivative term (\ref{totderivs}) since in the action we will have $\sqrt{-\hat{g}}\hat{\mathcal{L}}_{GB}\to\sqrt{-g}e^{[(p-3)\alpha+\beta n]\phi}(\mathcal{L}_{GB}+...)$. Taking this into account, we find
\beq
\begin{split}
\sqrt{-\hat{g}}\hat{\mathcal{L}}_{GB}&=e^{[(p-3)\alpha+\beta n]\phi}\biggr\{\mathcal{L}_{GB}+B_{1}G^{ab}(\nabla_{a}\phi)(\nabla_{b}\phi)+B_{2}\Box\phi(\nabla\phi)^{2}+B_{3}(\nabla\phi)^{4}\biggr\}\;,
\end{split}
\label{Ghatscaltenhighv2}\eeq
where 
\beq
\begin{split}
&B_{1}\equiv C_{3}-2C_{5}-C_{1}A= -4\left[(p-2)(p-3)\alpha^{2}+2n(p-2)\alpha\beta+n(n-1)\beta^{2}\right]\;,\\
&B_{2}\equiv C_{10}-\frac{1}{2}C_{6}+\frac{3}{2}AC_{7}=-2\biggr[(p-1)(p-2)(p-3)\alpha^{3}\\
&+3n(p-2)(p-1)\alpha^{2}\beta+3n(p-1)(n-1)\alpha\beta^{2}+n(n-1)(n-2)\beta^{3}\biggr]\;,\\
&B_{3}\equiv C_{9}-\frac{1}{2}C_{6}A+\frac{1}{2}A^{2}C_{7}=-(p-1)(p-2)^{2}(p-3)\alpha^{4}-4n(p-1)(p-2)^{2}\alpha^{3}\beta\\
&-2n(p-1)[3pn-2p-5n+3]\alpha^{2}\beta^{2}-4n(n-1)^{2}(p-1)\alpha\beta^{3}-n(n-1)^{2}(n-2)\beta^{4}\;,
\end{split}
\eeq
with $A\equiv(p-3)\alpha+n\beta$. The above is true up to total derivatives which we may now safely drop.

Since we will need it as well, we point out we can remove the $\Box\phi$ term appearing in the Ricci scalar (\ref{ricscal}) to arrive to 
\beq \sqrt{-\hat{g}}\hat{R}=e^{[(p-1)\alpha+n\beta]\phi}(R+\mathcal{A}(\nabla\phi)^{2})\;,\eeq
with
\beq \mathcal{A}\equiv A_{2}-[(p-1)\alpha+n\beta]A_{1}=p(p-1)\alpha^{2}+2np\alpha\beta+n(n-1)\beta^{2}\;.\eeq


\subsection{Equations of Motion}

Let us now compute the equations of motion for our theory, which we describe generically with the Lagrangian 
\beq \mathcal{L}=e^{A\phi}(R+a_{1}(\nabla\phi)^{2})-\frac{\alpha'}{4}e^{B\phi}\left[\mathcal{L}_{GB}+a_{2}G^{ab}(\nabla_{a}\phi)(\nabla_{b}\phi)+a_{3}\Box\phi(\nabla\phi)^{2}+a_{4}(\nabla\phi)^{4}\right]\;.\label{actiongen}\eeq
Here $A,B,a_{1},a_{2},a_{3}$, and $a_{4}$ are real coefficients, unrelated to the constants written above.

The gravitational field equations for a generic diffeomorphism invariant theory of gravity can be succinctly written as \cite{Padmanabhan:2007en}
\beq \bar{P}_{a}^{\;cde}R_{bcde}-2\nabla^{c}\nabla^{d}\bar{P}_{acdb}-\frac{1}{2}\mathcal{L}g_{ab}=8\pi GT_{ab}\;,\quad \bar{P}^{abcd}\equiv\frac{\partial\mathcal{L}}{\partial R_{abcd}}\;.\eeq
The tensor $\bar{P}^{abcd}$ manifestly shares the same algebraic symmetries as the Riemann tensor. For the action (\ref{actiongen}) we find\footnote{See, e.g., \cite{Svesko:2020yxo} for some useful formulas involving $\bar{P}^{abcd}$.}
\beq
\begin{split}
\bar{P}^{abcd}&=\frac{1}{2}e^{A\phi}(g^{ac}g^{bd}-g^{ad}g^{bc})\\
&-\frac{\alpha'}{4}e^{B\phi}\biggr\{(g_{ad}g_{bc}-g_{ab}g_{cd})R+\frac{a_{2}}{4}\biggr[g^{ac}(\nabla^{b}\phi)(\nabla^{d}\phi)+g^{bd}(\nabla^{a}\phi)(\nabla^{c}\phi)\\
&-g^{bc}(\nabla^{a}\phi)(\nabla^{d}\phi)-g^{ad}(\nabla^{b}\phi)(\nabla^{c}\phi)-(\nabla\phi)^{2}(g^{ac}g^{bd}-g^{ad}g^{bc})\biggr]\biggr\}\;,
\end{split}
\eeq

It is then straightforward to work out the gravitational field equations $\mathcal{E}_{ab}$:
\beq
\begin{split}
\mathcal{E}_{\mu\nu}&= e^{A\phi}\biggr(G_{\mu\nu}-b_{1}[A\nabla_{\mu}\nabla_{\nu}\phi+A^{2}(\nabla_{\mu}\phi)(\nabla_{\nu}\phi)]+g_{\mu\nu}[A\Box\phi+A^{2}(\nabla\phi)^{2}]\\
&+b_{1}(\nabla_{\mu}\phi)(\nabla_{\nu}\phi)-\frac{b_{1}}{2}g_{\mu\nu}(\nabla\phi)^{2}\biggr)\\
&+\frac{\alpha'}{8}e^{B\phi}\biggr\{g_{\mu\nu}b_{4}(\nabla\phi)^{4}+g_{\mu\nu}\mathcal{L}_{GB}+4R[B(\nabla_{\mu}\nabla_{\nu}\phi-g_{\mu\nu}\Box\phi)\\
&+B^{2}(\nabla_{\mu}\phi\nabla_{\nu}\phi-g_{\mu\nu}(\nabla\phi)^{2})]-[4RR_{\mu\nu}-8R_{\mu\gamma}R^{\gamma}_{\;\nu}-8R_{\mu\gamma\nu\delta}R^{\gamma\delta}+4R_{\mu\gamma\delta\rho}R_{\nu}^{\;\gamma\delta\rho}]\\
&+b_{2}\biggr[G_{\gamma\delta}(\nabla^{\gamma}\phi)(\nabla^{\delta}\phi)+2R_{\mu\delta\gamma\nu}(\nabla^{\gamma}\phi)(\nabla^{\delta}\phi)+g_{\mu\nu}R_{\gamma\delta}(\nabla^{\gamma}\phi)(\nabla^{\delta}\phi)\\
&-R_{\nu\delta}(\nabla_{\mu}\phi)(\nabla^{\delta}\phi)+R_{\mu\nu}(\nabla\phi)^{2}+2(\nabla_{\mu}\nabla_{\nu}\phi)\Box\phi-2(\nabla_{\mu}\nabla^{\gamma}\phi)(\nabla_{\nu}\nabla_{\gamma}\phi)+g_{\mu\nu}[(\nabla\nabla\phi)^{2}-(\Box\phi)^{2}]\\
&+B\biggr((\nabla_{\gamma}\phi)(\nabla_{\mu}\phi)(\nabla_{\nu}\nabla^{\gamma}\phi)+(\nabla_{\gamma}\phi)(\nabla_{\nu}\phi)(\nabla_{\mu}\nabla^{\gamma}\phi)-\Box\phi(\nabla_{\mu}\phi)(\nabla_{\nu}\phi)\\
&-g_{\mu\nu}(\nabla\nabla\phi)(\nabla\phi)(\nabla\phi)+(\nabla\phi)^{2}[g_{\mu\nu}\Box\phi-\nabla_{\mu}\nabla_{\nu}\phi]\biggr)\biggr]-4b_{4}(\nabla\phi)^{2}(\nabla_{\mu}\phi)(\nabla_{\nu}\phi)\\
&-b_{3}\biggr[g_{\mu\nu}[2(\nabla_{\alpha}\phi)(\nabla_{\beta}\phi)(\nabla^{\alpha}\nabla^{\beta}\phi)+B(\nabla\phi)^{4}]+2\Box\phi(\nabla_{\mu}\phi)(\nabla_{\nu}\phi)\\
&-2(\nabla^{\alpha}\phi)(\nabla_{\nu}\phi)(\nabla_{\alpha}\nabla_{\mu}\phi)-2(\nabla^{\alpha}\phi)(\nabla_{\mu}\phi)(\nabla_{\alpha}\nabla_{\nu}\phi)-2B(\nabla\phi)^{2}(\nabla_{\mu}\phi)(\nabla_{\nu}\phi)\biggr]\biggr\}\;,
\end{split}
\label{graveom}\eeq

Meanwhile the equations of motion for the dilaton are easily worked out to be
\beq
\begin{split}
\mathcal{E}_{\phi}&=e^{A\phi}[AR-a_{1}A(\nabla\phi)^{2}-2a_{1}\Box\phi]\\
&-\frac{\alpha'}{4}e^{B\phi}\biggr\{B\mathcal{L}_{GB}+(2R^{ab}-a_{2}BG^{ab})(\nabla_{a}\phi)(\nabla_{b}\phi)-2a_{2}G^{ab}(\nabla_{a}\nabla_{b}\phi)\\
&+(a_{3}B^{2}-3a_{4}B)(\nabla\phi)^{4}+(6a_{3}B-8a_{4})(\nabla\nabla\phi)(\nabla\phi)(\nabla\phi)-2a_{3}(\Box\phi)^{2}-4a_{4}\Box\phi(\nabla\phi)^{2}\biggr\}\;.
\end{split}
\label{phieom}\eeq


\bibliography{referencesGR}

\providecommand{\href}[2]{#2}\begingroup\raggedright\begin{thebibliography}{100}

\bibitem{Clifton:2011jh}
T.~Clifton, P.~G. Ferreira, A.~Padilla and C.~Skordis, \emph{{Modified Gravity
  and Cosmology}},
  \href{http://dx.doi.org/10.1016/j.physrep.2012.01.001}{\emph{Phys. Rept.}
  {\bf 513} (2012) 1--189}, [\href{https://arxiv.org/abs/1106.2476}{{\tt
  1106.2476}}].

\bibitem{Lovelock:1971yv}
D.~Lovelock, \emph{{The Einstein tensor and its generalizations}},
  \href{http://dx.doi.org/10.1063/1.1665613}{\emph{J. Math. Phys.} {\bf 12}
  (1971) 498--501}.

\bibitem{Lovelock:1972vz}
D.~Lovelock, \emph{{The four-dimensionality of space and the einstein tensor}},
  \href{http://dx.doi.org/10.1063/1.1666069}{\emph{J. Math. Phys.} {\bf 13}
  (1972) 874--876}.

\bibitem{Teyssandier:1983zz}
P.~Teyssandier and P.~Tourrenc, \emph{{The Cauchy problem for the R+R**2
  theories of gravity without torsion}},
  \href{http://dx.doi.org/10.1063/1.525659}{\emph{J. Math. Phys.} {\bf 24}
  (1983) 2793}.

\bibitem{Maeda:1988ab}
K.-i. Maeda, \emph{{Towards the Einstein-Hilbert Action via Conformal
  Transformation}},
  \href{http://dx.doi.org/10.1103/PhysRevD.39.3159}{\emph{Phys. Rev. D} {\bf
  39} (1989) 3159}.

\bibitem{Wands:1993uu}
D.~Wands, \emph{{Extended gravity theories and the Einstein-Hilbert action}},
  \href{http://dx.doi.org/10.1088/0264-9381/11/1/025}{\emph{Class. Quant.
  Grav.} {\bf 11} (1994) 269--280},
  [\href{https://arxiv.org/abs/gr-qc/9307034}{{\tt gr-qc/9307034}}].

\bibitem{Magnano:1993bd}
G.~Magnano and L.~M. Sokolowski, \emph{{On physical equivalence between
  nonlinear gravity theories and a general relativistic selfgravitating scalar
  field}}, \href{http://dx.doi.org/10.1103/PhysRevD.50.5039}{\emph{Phys. Rev.
  D} {\bf 50} (1994) 5039--5059},
  [\href{https://arxiv.org/abs/gr-qc/9312008}{{\tt gr-qc/9312008}}].

\bibitem{Horndeski74-1}
G.~W. Horndeski, \emph{{Second-order scalar-tensor field equations in a
  four-dimensional space}},
  \href{http://dx.doi.org/10.1007/BF01807638}{\emph{Int. J. Theor. Phys.} {\bf
  10} (1974) 363--384}.

\bibitem{Deffayet:2009wt}
C.~Deffayet, G.~Esposito-Farese and A.~Vikman, \emph{{Covariant Galileon}},
  \href{http://dx.doi.org/10.1103/PhysRevD.79.084003}{\emph{Phys. Rev. D} {\bf
  79} (2009) 084003}, [\href{https://arxiv.org/abs/0901.1314}{{\tt
  0901.1314}}].

\bibitem{Deffayet:2009mn}
C.~Deffayet, S.~Deser and G.~Esposito-Farese, \emph{{Generalized Galileons: All
  scalar models whose curved background extensions maintain second-order field
  equations and stress-tensors}},
  \href{http://dx.doi.org/10.1103/PhysRevD.80.064015}{\emph{Phys. Rev. D} {\bf
  80} (2009) 064015}, [\href{https://arxiv.org/abs/0906.1967}{{\tt
  0906.1967}}].

\bibitem{Deffayet:2011gz}
C.~Deffayet, X.~Gao, D.~Steer and G.~Zahariade, \emph{{From k-essence to
  generalised Galileons}},
  \href{http://dx.doi.org/10.1103/PhysRevD.84.064039}{\emph{Phys. Rev. D} {\bf
  84} (2011) 064039}, [\href{https://arxiv.org/abs/1103.3260}{{\tt
  1103.3260}}].

\bibitem{Kobayashi:2011nu}
T.~Kobayashi, M.~Yamaguchi and J.~Yokoyama, \emph{{Generalized G-inflation:
  Inflation with the most general second-order field equations}},
  \href{http://dx.doi.org/10.1143/PTP.126.511}{\emph{Prog. Theor. Phys.} {\bf
  126} (2011) 511--529}, [\href{https://arxiv.org/abs/1105.5723}{{\tt
  1105.5723}}].

\bibitem{Charmousis14-1}
C.~Charmousis, \emph{{From Lovelock to Horndeski`s Generalized Scalar Tensor
  Theory}}, \href{http://dx.doi.org/10.1007/978-3-319-10070-8_2}{\emph{Lect.
  Notes Phys.} {\bf 892} (2015) 25--56},
  [\href{https://arxiv.org/abs/1405.1612}{{\tt 1405.1612}}].

\bibitem{Brans:1961sx}
C.~Brans and R.~Dicke, \emph{{Mach's principle and a relativistic theory of
  gravitation}}, \href{http://dx.doi.org/10.1103/PhysRev.124.925}{\emph{Phys.
  Rev.} {\bf 124} (1961) 925--935}.

\bibitem{ArmendarizPicon:1999rj}
C.~Armendariz-Picon, T.~Damour and V.~F. Mukhanov, \emph{{k - inflation}},
  \href{http://dx.doi.org/10.1016/S0370-2693(99)00603-6}{\emph{Phys. Lett. B}
  {\bf 458} (1999) 209--218}, [\href{https://arxiv.org/abs/hep-th/9904075}{{\tt
  hep-th/9904075}}].

\bibitem{ArmendarizPicon:2000dh}
C.~Armendariz-Picon, V.~F. Mukhanov and P.~J. Steinhardt, \emph{{A Dynamical
  solution to the problem of a small cosmological constant and late time cosmic
  acceleration}},
  \href{http://dx.doi.org/10.1103/PhysRevLett.85.4438}{\emph{Phys. Rev. Lett.}
  {\bf 85} (2000) 4438--4441},
  [\href{https://arxiv.org/abs/astro-ph/0004134}{{\tt astro-ph/0004134}}].

\bibitem{ArmendarizPicon:2000ah}
C.~Armendariz-Picon, V.~F. Mukhanov and P.~J. Steinhardt, \emph{{Essentials of
  k essence}}, \href{http://dx.doi.org/10.1103/PhysRevD.63.103510}{\emph{Phys.
  Rev. D} {\bf 63} (2001) 103510},
  [\href{https://arxiv.org/abs/astro-ph/0006373}{{\tt astro-ph/0006373}}].

\bibitem{Charmousis:2011bf}
C.~Charmousis, E.~J. Copeland, A.~Padilla and P.~M. Saffin, \emph{{General
  second order scalar-tensor theory, self tuning, and the Fab Four}},
  \href{http://dx.doi.org/10.1103/PhysRevLett.108.051101}{\emph{Phys. Rev.
  Lett.} {\bf 108} (2012) 051101}, [\href{https://arxiv.org/abs/1106.2000}{{\tt
  1106.2000}}].

\bibitem{Charmousis:2011ea}
C.~Charmousis, E.~J. Copeland, A.~Padilla and P.~M. Saffin, \emph{{Self-tuning
  and the derivation of a class of scalar-tensor theories}},
  \href{http://dx.doi.org/10.1103/PhysRevD.85.104040}{\emph{Phys. Rev. D} {\bf
  85} (2012) 104040}, [\href{https://arxiv.org/abs/1112.4866}{{\tt
  1112.4866}}].

\bibitem{Nicolis:2008in}
A.~Nicolis, R.~Rattazzi and E.~Trincherini, \emph{{The Galileon as a local
  modification of gravity}},
  \href{http://dx.doi.org/10.1103/PhysRevD.79.064036}{\emph{Phys. Rev. D} {\bf
  79} (2009) 064036}, [\href{https://arxiv.org/abs/0811.2197}{{\tt
  0811.2197}}].

\bibitem{Silva:2009km}
F.~P. Silva and K.~Koyama, \emph{{Self-Accelerating Universe in Galileon
  Cosmology}}, \href{http://dx.doi.org/10.1103/PhysRevD.80.121301}{\emph{Phys.
  Rev. D} {\bf 80} (2009) 121301}, [\href{https://arxiv.org/abs/0909.4538}{{\tt
  0909.4538}}].

\bibitem{Creminelli10-1}
P.~Creminelli, A.~Nicolis and E.~Trincherini, \emph{{Galilean Genesis: An
  Alternative to inflation}},
  \href{http://dx.doi.org/10.1088/1475-7516/2010/11/021}{\emph{JCAP} {\bf 1011}
  (2010) 021}, [\href{https://arxiv.org/abs/1007.0027}{{\tt 1007.0027}}].

\bibitem{Kobayashi:2010cm}
T.~Kobayashi, M.~Yamaguchi and J.~Yokoyama, \emph{{G-inflation: Inflation
  driven by the Galileon field}},
  \href{http://dx.doi.org/10.1103/PhysRevLett.105.231302}{\emph{Phys. Rev.
  Lett.} {\bf 105} (2010) 231302}, [\href{https://arxiv.org/abs/1008.0603}{{\tt
  1008.0603}}].

\bibitem{Deffayet:2010qz}
C.~Deffayet, O.~Pujolas, I.~Sawicki and A.~Vikman, \emph{{Imperfect Dark Energy
  from Kinetic Gravity Braiding}},
  \href{http://dx.doi.org/10.1088/1475-7516/2010/10/026}{\emph{JCAP} {\bf 10}
  (2010) 026}, [\href{https://arxiv.org/abs/1008.0048}{{\tt 1008.0048}}].

\bibitem{Qiu:2011cy}
T.~Qiu, J.~Evslin, Y.-F. Cai, M.~Li and X.~Zhang, \emph{{Bouncing Galileon
  Cosmologies}},
  \href{http://dx.doi.org/10.1088/1475-7516/2011/10/036}{\emph{JCAP} {\bf 10}
  (2011) 036}, [\href{https://arxiv.org/abs/1108.0593}{{\tt 1108.0593}}].

\bibitem{Easson:2011zy}
D.~A. Easson, I.~Sawicki and A.~Vikman, \emph{{G-Bounce}},
  \href{http://dx.doi.org/10.1088/1475-7516/2011/11/021}{\emph{JCAP} {\bf 11}
  (2011) 021}, [\href{https://arxiv.org/abs/1109.1047}{{\tt 1109.1047}}].

\bibitem{Rubakov13-1}
V.~A. Rubakov, \emph{{Consistent NEC-violation: towards creating a universe in
  the laboratory}},
  \href{http://dx.doi.org/10.1103/PhysRevD.88.044015}{\emph{Phys. Rev.} {\bf
  D88} (2013) 044015}, [\href{https://arxiv.org/abs/1305.2614}{{\tt
  1305.2614}}].

\bibitem{Ijjas16-1}
A.~Ijjas and P.~J. Steinhardt, \emph{{Fully stable cosmological solutions with
  a non-singular classical bounce}},
  \href{http://dx.doi.org/10.1016/j.physletb.2016.11.047}{\emph{Phys. Lett.}
  {\bf B764} (2017) 289--294}, [\href{https://arxiv.org/abs/1609.01253}{{\tt
  1609.01253}}].

\bibitem{VanAcoleyen11-1}
K.~Van~Acoleyen and J.~Van~Doorsselaere, \emph{{Galileons from Lovelock
  actions}}, \href{http://dx.doi.org/10.1103/PhysRevD.83.084025}{\emph{Phys.
  Rev.} {\bf D83} (2011) 084025}, [\href{https://arxiv.org/abs/1102.0487}{{\tt
  1102.0487}}].

\bibitem{Zwiebach85-1}
B.~Zwiebach, \emph{Curvature squared terms in string theories}, {\emph{Phys.
  Lett. B} {\bf 156} (1985) }.

\bibitem{Sen:1985qt}
A.~Sen, \emph{{Equations of Motion for the Heterotic String Theory from the
  Conformal Invariance of the Sigma Model}},
  \href{http://dx.doi.org/10.1103/PhysRevLett.55.1846}{\emph{Phys. Rev. Lett.}
  {\bf 55} (1985) 1846}.

\bibitem{Gross86-1}
D.~J. Gross and E.~Witten, \emph{Superstring modifications of einstein's
  equations}, {\emph{Nucl. Phys. B} {\bf 277} (1986) }.

\bibitem{Gross:1986mw}
D.~J. Gross and J.~H. Sloan, \emph{{The Quartic Effective Action for the
  Heterotic String}},
  \href{http://dx.doi.org/10.1016/0550-3213(87)90465-2}{\emph{Nucl. Phys. B}
  {\bf 291} (1987) 41--89}.

\bibitem{Metsaev:1987zx}
R.~Metsaev and A.~A. Tseytlin, \emph{{Order alpha-prime (Two Loop) Equivalence
  of the String Equations of Motion and the Sigma Model Weyl Invariance
  Conditions: Dependence on the Dilaton and the Antisymmetric Tensor}},
  \href{http://dx.doi.org/10.1016/0550-3213(87)90077-0}{\emph{Nucl. Phys. B}
  {\bf 293} (1987) 385--419}.

\bibitem{Charmousis12-1}
C.~Charmousis, B.~Gouteraux and E.~Kiritsis, \emph{{Higher-derivative
  scalar-vector-tensor theories: black holes, Galileons, singularity cloaking
  and holography}},
  \href{http://dx.doi.org/10.1007/JHEP09(2012)011}{\emph{JHEP} {\bf 09} (2012)
  011}, [\href{https://arxiv.org/abs/1206.1499}{{\tt 1206.1499}}].

\bibitem{Glavan:2019inb}
D.~z. Glavan and C.~Lin, \emph{{Einstein-Gauss-Bonnet gravity in 4-dimensional
  space-time}},
  \href{http://dx.doi.org/10.1103/PhysRevLett.124.081301}{\emph{Phys.\ Rev.\
  Lett.} {\bf 124} (2020) 081301},
  [\href{https://arxiv.org/abs/1905.03601}{{\tt 1905.03601}}].

\bibitem{Konoplya:2020qqh}
R.~Konoplya and A.~Zhidenko, \emph{{Black holes in the four-dimensional
  Einstein-Lovelock gravity}},
  \href{http://dx.doi.org/10.1103/PhysRevD.101.084038}{\emph{Phys. Rev. D} {\bf
  101} (2020) 084038}, [\href{https://arxiv.org/abs/2003.07788}{{\tt
  2003.07788}}].

\bibitem{Kumar:2020owy}
R.~Kumar and S.~G. Ghosh, \emph{{Rotating black holes in the novel $4D$
  Einstein-Gauss-Bonnet gravity}},
  \href{https://arxiv.org/abs/2003.08927}{{\tt 2003.08927}}.

\bibitem{Mishra:2020gce}
A.~K. Mishra, \emph{{Quasinormal modes and Strong Cosmic Censorship in the
  novel 4D Einstein-Gauss-Bonnet gravity}},
  \href{https://arxiv.org/abs/2004.01243}{{\tt 2004.01243}}.

\bibitem{Doneva:2020ped}
D.~D. Doneva and S.~S. Yazadjiev, \emph{{Relativistic stars in 4D
  Einstein-Gauss-Bonnet gravity}},
  \href{https://arxiv.org/abs/2003.10284}{{\tt 2003.10284}}.

\bibitem{Odintsov:2020sqy}
S.~Odintsov, V.~Oikonomou and F.~Fronimos, \emph{{Rectifying
  Einstein-Gauss-Bonnet Inflation in View of GW170817}},
  \href{https://arxiv.org/abs/2003.13724}{{\tt 2003.13724}}.

\bibitem{Jusufi:2020yus}
K.~Jusufi, A.~Banerjee and S.~G. Ghosh, \emph{{Wormholes in 4D
  Einstein-Gauss-Bonnet Gravity}},
  \href{https://arxiv.org/abs/2004.10750}{{\tt 2004.10750}}.

\bibitem{Arrechea:2020evj}
J.~Arrechea, A.~Delhom and A.~Jiménez-Cano, \emph{{Yet another comment on
  four-dimensional Einstein-Gauss-Bonnet gravity}},
  \href{https://arxiv.org/abs/2004.12998}{{\tt 2004.12998}}.

\bibitem{Aoki:2020lig}
K.~Aoki, M.~A. Gorji and S.~Mukohyama, \emph{{A consistent theory of
  $D\rightarrow 4$ Einstein-Gauss-Bonnet gravity}},
  \href{https://arxiv.org/abs/2005.03859}{{\tt 2005.03859}}.

\bibitem{Lu:2020iav}
H.~Lu and Y.~Pang, \emph{{Horndeski Gravity as $D\rightarrow4$ Limit of
  Gauss-Bonnet}},  \href{https://arxiv.org/abs/2003.11552}{{\tt 2003.11552}}.

\bibitem{Gurses:2020ofy}
M.~Gurses, T.~C. Sisman and B.~Tekin, \emph{{Is there a novel
  Einstein-Gauss-Bonnet theory in four dimensions?}},
  \href{https://arxiv.org/abs/2004.03390}{{\tt 2004.03390}}.

\bibitem{Ai:2020peo}
W.-Y. Ai, \emph{{A note on the novel 4D Einstein-Gauss-Bonnet gravity}},
  \href{https://arxiv.org/abs/2004.02858}{{\tt 2004.02858}}.

\bibitem{Mahapatra:2020rds}
S.~Mahapatra, \emph{{A note on the total action of $4D$ Gauss-Bonnet theory}},
  \href{https://arxiv.org/abs/2004.09214}{{\tt 2004.09214}}.

\bibitem{Fernandes:2020nbq}
P.~G. Fernandes, P.~Carrilho, T.~Clifton and D.~J. Mulryne, \emph{{Derivation
  of Regularized Field Equations for the Einstein-Gauss-Bonnet Theory in Four
  Dimensions}},  \href{https://arxiv.org/abs/2004.08362}{{\tt 2004.08362}}.

\bibitem{Hennigar:2020lsl}
R.~A. Hennigar, D.~Kubiznak, R.~B. Mann and C.~Pollack, \emph{{On Taking the
  $D\to 4$ limit of Gauss-Bonnet Gravity: Theory and Solutions}},
  \href{https://arxiv.org/abs/2004.09472}{{\tt 2004.09472}}.

\bibitem{Easson:2020mpq}
D.~A. Easson, T.~Manton and A.~Svesko, \emph{{$D\to4$ Einstein-Gauss-Bonnet
  Gravity and Beyond}},
  \href{http://dx.doi.org/10.1088/1475-7516/2020/10/026}{\emph{JCAP} {\bf 10}
  (2020) 026}, [\href{https://arxiv.org/abs/2005.12292}{{\tt 2005.12292}}].

\bibitem{Charmousis:2014mia}
C.~Charmousis, \emph{{From Lovelock to Horndeski`s Generalized Scalar Tensor
  Theory}}, \href{http://dx.doi.org/10.1007/978-3-319-10070-8\_2}{\emph{Lect.\
  Notes Phys.} {\bf 892} (2015) 25--56},
  [\href{https://arxiv.org/abs/1405.1612}{{\tt 1405.1612}}].

\bibitem{Mann:1992ar}
R.~B. Mann and S.~Ross, \emph{{The D $\to$ 2 limit of general relativity}},
  \href{http://dx.doi.org/10.1088/0264-9381/10/7/015}{\emph{Class.\ Quant.\
  Grav.} {\bf 10} (1993) 1405--1408},
  [\href{https://arxiv.org/abs/gr-qc/9208004}{{\tt gr-qc/9208004}}].

\bibitem{MuellerHoissen:1989yv}
F.~Mueller-Hoissen, \emph{{Gravity Actions, Boundary Terms and Second Order
  Field Equations}},
  \href{http://dx.doi.org/10.1016/0550-3213(90)90513-D}{\emph{Nucl. Phys. B}
  {\bf 337} (1990) 709--736}.

\bibitem{Duff86-1}
M.~J. Duff, B.~E.~W. Nilsson and C.~N. Pope, \emph{{Kaluza-Klein
  Supergravity}},
  \href{http://dx.doi.org/10.1016/0370-1573(86)90163-8}{\emph{Phys. Rept.} {\bf
  130} (1986) 1--142}.

\bibitem{Cvetic00-1}
M.~Cvetic, H.~Lu and C.~N. Pope, \emph{{Consistent Kaluza-Klein sphere
  reductions}}, \href{http://dx.doi.org/10.1103/PhysRevD.62.064028}{\emph{Phys.
  Rev.} {\bf D62} (2000) 064028},
  [\href{https://arxiv.org/abs/hep-th/0003286}{{\tt hep-th/0003286}}].

\bibitem{Gouteraux:2011qh}
B.~Gouteraux, J.~Smolic, M.~Smolic, K.~Skenderis and M.~Taylor,
  \emph{{Holography for Einstein-Maxwell-dilaton theories from generalized
  dimensional reduction}},
  \href{http://dx.doi.org/10.1007/JHEP01(2012)089}{\emph{JHEP} {\bf 01} (2012)
  089}, [\href{https://arxiv.org/abs/1110.2320}{{\tt 1110.2320}}].

\bibitem{Kanitscheider:2009as}
I.~Kanitscheider and K.~Skenderis, \emph{{Universal hydrodynamics of
  non-conformal branes}},
  \href{http://dx.doi.org/10.1088/1126-6708/2009/04/062}{\emph{JHEP} {\bf 04}
  (2009) 062}, [\href{https://arxiv.org/abs/0901.1487}{{\tt 0901.1487}}].

\bibitem{Liu:2020yhu}
P.~Liu, C.~Niu, X.~Wang and C.-Y. Zhang, \emph{{Traversable Thin-shell Wormhole
  in the Novel 4D Einstein-Gauss-Bonnet Theory}},
  \href{https://arxiv.org/abs/2004.14267}{{\tt 2004.14267}}.

\bibitem{Gasperini07-1}
M.~Gasperini, \emph{{Elements of string cosmology}}.
\newblock Cambridge University Press, 2007.

\bibitem{Deffayet:2010zh}
C.~Deffayet, S.~Deser and G.~Esposito-Farese, \emph{{Arbitrary $p$-form
  Galileons}}, \href{http://dx.doi.org/10.1103/PhysRevD.82.061501}{\emph{Phys.
  Rev. D} {\bf 82} (2010) 061501}, [\href{https://arxiv.org/abs/1007.5278}{{\tt
  1007.5278}}].

\bibitem{Maeda:2011zn}
K.-i. Maeda, N.~Ohta and R.~Wakebe, \emph{{Accelerating Universes in String
  Theory via Field Redefinition}},
  \href{http://dx.doi.org/10.1140/epjc/s10052-012-1949-6}{\emph{Eur. Phys. J.
  C} {\bf 72} (2012) 1949}, [\href{https://arxiv.org/abs/1111.3251}{{\tt
  1111.3251}}].

\bibitem{Hui:2012qt}
L.~Hui and A.~Nicolis, \emph{{No-Hair Theorem for the Galileon}},
  \href{http://dx.doi.org/10.1103/PhysRevLett.110.241104}{\emph{Phys. Rev.
  Lett.} {\bf 110} (2013) 241104}, [\href{https://arxiv.org/abs/1202.1296}{{\tt
  1202.1296}}].

\bibitem{Babichev:2016rlq}
E.~Babichev, C.~Charmousis and A.~Leh\'ebel, \emph{{Black holes and stars in
  Horndeski theory}},
  \href{http://dx.doi.org/10.1088/0264-9381/33/15/154002}{\emph{Class. Quant.
  Grav.} {\bf 33} (2016) 154002}, [\href{https://arxiv.org/abs/1604.06402}{{\tt
  1604.06402}}].

\bibitem{Weinberg:1988cp}
S.~Weinberg, \emph{{The Cosmological Constant Problem}},
  \href{http://dx.doi.org/10.1103/RevModPhys.61.1}{\emph{Rev. Mod. Phys.} {\bf
  61} (1989) 1--23}.

\bibitem{Vainshtein:1972sx}
A.~I. Vainshtein, \emph{{To the problem of nonvanishing gravitation mass}},
  \href{http://dx.doi.org/10.1016/0370-2693(72)90147-5}{\emph{Phys. Lett. B}
  {\bf 39} (1972) 393--394}.

\bibitem{Babichev:2013usa}
E.~Babichev and C.~Deffayet, \emph{{An introduction to the Vainshtein
  mechanism}},
  \href{http://dx.doi.org/10.1088/0264-9381/30/18/184001}{\emph{Class. Quant.
  Grav.} {\bf 30} (2013) 184001}, [\href{https://arxiv.org/abs/1304.7240}{{\tt
  1304.7240}}].

\bibitem{Babichev:2009ee}
E.~Babichev, C.~Deffayet and R.~Ziour, \emph{{k-Mouflage gravity}},
  \href{http://dx.doi.org/10.1142/S0218271809016107}{\emph{Int. J. Mod. Phys.
  D} {\bf 18} (2009) 2147--2154}, [\href{https://arxiv.org/abs/0905.2943}{{\tt
  0905.2943}}].

\bibitem{Callan:1992rs}
J.~Callan, Curtis~G., S.~B. Giddings, J.~A. Harvey and A.~Strominger,
  \emph{{Evanescent black holes}},
  \href{http://dx.doi.org/10.1103/PhysRevD.45.R1005}{\emph{Phys. Rev. D} {\bf
  45} (1992) 1005}, [\href{https://arxiv.org/abs/hep-th/9111056}{{\tt
  hep-th/9111056}}].

\bibitem{Rosso:2020zkk}
F.~Rosso and A.~Svesko, \emph{{Novel aspects of the extended first law of
  entanglement}}, \href{http://dx.doi.org/10.1007/JHEP08(2020)008}{\emph{JHEP}
  {\bf 08} (2020) 008}, [\href{https://arxiv.org/abs/2003.10462}{{\tt
  2003.10462}}].

\bibitem{Sawicki:2012pz}
I.~Sawicki and A.~Vikman, \emph{{Hidden Negative Energies in Strongly
  Accelerated Universes}},
  \href{http://dx.doi.org/10.1103/PhysRevD.87.067301}{\emph{Phys. Rev. D} {\bf
  87} (2013) 067301}, [\href{https://arxiv.org/abs/1209.2961}{{\tt
  1209.2961}}].

\bibitem{Parikh:2015ret}
M.~Parikh and A.~Svesko, \emph{{Thermodynamic Origin of the Null Energy
  Condition}}, \href{http://dx.doi.org/10.1103/PhysRevD.95.104002}{\emph{Phys.
  Rev. D} {\bf 95} (2017) 104002},
  [\href{https://arxiv.org/abs/1511.06460}{{\tt 1511.06460}}].

\bibitem{Parikh:2016lys}
M.~Parikh and A.~Svesko, \emph{{Logarithmic corrections to gravitational
  entropy and the null energy condition}},
  \href{http://dx.doi.org/10.1016/j.physletb.2016.07.071}{\emph{Phys. Lett. B}
  {\bf 761} (2016) 16--19}, [\href{https://arxiv.org/abs/1612.06949}{{\tt
  1612.06949}}].

\bibitem{Parikh:2014mja}
M.~Parikh and J.~P. van~der Schaar, \emph{{Derivation of the Null Energy
  Condition}}, \href{http://dx.doi.org/10.1103/PhysRevD.91.084002}{\emph{Phys.
  Rev. D} {\bf 91} (2015) 084002}, [\href{https://arxiv.org/abs/1406.5163}{{\tt
  1406.5163}}].

\bibitem{Brandenberger:2009jq}
R.~H. Brandenberger, \emph{{Alternatives to the inflationary paradigm of
  structure formation}},
  \href{http://dx.doi.org/10.1142/S2010194511000109}{\emph{Int. J. Mod. Phys.
  Conf. Ser.} {\bf 01} (2011) 67--79},
  [\href{https://arxiv.org/abs/0902.4731}{{\tt 0902.4731}}].

\bibitem{Chatterjee:2012zh}
S.~Chatterjee, D.~A. Easson and M.~Parikh, \emph{{Energy conditions in the
  Jordan frame}},
  \href{http://dx.doi.org/10.1088/0264-9381/30/23/235031}{\emph{Class. Quant.
  Grav.} {\bf 30} (2013) 235031}, [\href{https://arxiv.org/abs/1212.6430}{{\tt
  1212.6430}}].

\bibitem{Adams:2006sv}
A.~Adams, N.~Arkani-Hamed, S.~Dubovsky, A.~Nicolis and R.~Rattazzi,
  \emph{{Causality, analyticity and an IR obstruction to UV completion}},
  \href{http://dx.doi.org/10.1088/1126-6708/2006/10/014}{\emph{JHEP} {\bf 10}
  (2006) 014}, [\href{https://arxiv.org/abs/hep-th/0602178}{{\tt
  hep-th/0602178}}].

\bibitem{Shore:2007um}
G.~M. Shore, \emph{{Superluminality and UV completion}},
  \href{http://dx.doi.org/10.1016/j.nuclphysb.2007.03.034}{\emph{Nucl. Phys. B}
  {\bf 778} (2007) 219--258}, [\href{https://arxiv.org/abs/hep-th/0701185}{{\tt
  hep-th/0701185}}].

\bibitem{Easson:2013bda}
D.~A. Easson, I.~Sawicki and A.~Vikman, \emph{{When Matter Matters}},
  \href{http://dx.doi.org/10.1088/1475-7516/2013/07/014}{\emph{JCAP} {\bf 07}
  (2013) 014}, [\href{https://arxiv.org/abs/1304.3903}{{\tt 1304.3903}}].

\bibitem{Dobre:2017pnt}
D.~A. Dobre, A.~V. Frolov, J.~T. G\'alvez~Ghersi, S.~Ramazanov and A.~Vikman,
  \emph{{Unbraiding the Bounce: Superluminality around the Corner}},
  \href{http://dx.doi.org/10.1088/1475-7516/2018/03/020}{\emph{JCAP} {\bf 03}
  (2018) 020}, [\href{https://arxiv.org/abs/1712.10272}{{\tt 1712.10272}}].

\bibitem{deRham:2021fpu}
C.~de~Rham, S.~Melville and J.~Noller, \emph{{Positivity Bounds on Dark Energy:
  When Matter Matters}},  \href{https://arxiv.org/abs/2103.06855}{{\tt
  2103.06855}}.

\bibitem{Dvali:2010jz}
G.~Dvali, G.~F. Giudice, C.~Gomez and A.~Kehagias, \emph{{UV-Completion by
  Classicalization}},
  \href{http://dx.doi.org/10.1007/JHEP08(2011)108}{\emph{JHEP} {\bf 08} (2011)
  108}, [\href{https://arxiv.org/abs/1010.1415}{{\tt 1010.1415}}].

\bibitem{Dvali:2010ns}
G.~Dvali and D.~Pirtskhalava, \emph{{Dynamics of Unitarization by
  Classicalization}},
  \href{http://dx.doi.org/10.1016/j.physletb.2011.03.054}{\emph{Phys. Lett. B}
  {\bf 699} (2011) 78--86}, [\href{https://arxiv.org/abs/1011.0114}{{\tt
  1011.0114}}].

\bibitem{Vafa:2005ui}
C.~Vafa, \emph{{The String landscape and the swampland}},
  \href{https://arxiv.org/abs/hep-th/0509212}{{\tt hep-th/0509212}}.

\bibitem{Ooguri:2006in}
H.~Ooguri and C.~Vafa, \emph{{On the Geometry of the String Landscape and the
  Swampland}},
  \href{http://dx.doi.org/10.1016/j.nuclphysb.2006.10.033}{\emph{Nucl. Phys. B}
  {\bf 766} (2007) 21--33}, [\href{https://arxiv.org/abs/hep-th/0605264}{{\tt
  hep-th/0605264}}].

\bibitem{Palti:2019pca}
E.~Palti, \emph{{The Swampland: Introduction and Review}},
  \href{http://dx.doi.org/10.1002/prop.201900037}{\emph{Fortsch. Phys.} {\bf
  67} (2019) 1900037}, [\href{https://arxiv.org/abs/1903.06239}{{\tt
  1903.06239}}].

\bibitem{vanBeest:2021lhn}
M.~van Beest, J.~Calder\'on-Infante, D.~Mirfendereski and I.~Valenzuela,
  \emph{{Lectures on the Swampland Program in String Compactifications}},
  \href{https://arxiv.org/abs/2102.01111}{{\tt 2102.01111}}.

\bibitem{Heisenberg:2019qxz}
L.~Heisenberg, M.~Bartelmann, R.~Brandenberger and A.~Refregier,
  \emph{{Horndeski gravity in the swampland}},
  \href{http://dx.doi.org/10.1103/PhysRevD.99.124020}{\emph{Phys. Rev. D} {\bf
  99} (2019) 124020}, [\href{https://arxiv.org/abs/1902.03939}{{\tt
  1902.03939}}].

\bibitem{Brahma:2019kch}
S.~Brahma and M.~W. Hossain, \emph{{Dark energy beyond quintessence:
  Constraints from the swampland}},
  \href{http://dx.doi.org/10.1007/JHEP06(2019)070}{\emph{JHEP} {\bf 06} (2019)
  070}, [\href{https://arxiv.org/abs/1902.11014}{{\tt 1902.11014}}].

\bibitem{Agrawal:2018own}
P.~Agrawal, G.~Obied, P.~J. Steinhardt and C.~Vafa, \emph{{On the Cosmological
  Implications of the String Swampland}},
  \href{http://dx.doi.org/10.1016/j.physletb.2018.07.040}{\emph{Phys. Lett. B}
  {\bf 784} (2018) 271--276}, [\href{https://arxiv.org/abs/1806.09718}{{\tt
  1806.09718}}].

\bibitem{Obied:2018sgi}
G.~Obied, H.~Ooguri, L.~Spodyneiko and C.~Vafa, \emph{{De Sitter Space and the
  Swampland}},  \href{https://arxiv.org/abs/1806.08362}{{\tt 1806.08362}}.

\bibitem{Wang:2020eln}
P.~Wang, H.~Wu, H.~Yang and S.~Ying, \emph{{Derive Lovelock Gravity from String
  Theory in Cosmological Background}},
  \href{https://arxiv.org/abs/2012.13312}{{\tt 2012.13312}}.

\bibitem{Padilla:2010de}
A.~Padilla, P.~M. Saffin and S.-Y. Zhou, \emph{{Bi-galileon theory I:
  Motivation and formulation}},
  \href{http://dx.doi.org/10.1007/JHEP12(2010)031}{\emph{JHEP} {\bf 12} (2010)
  031}, [\href{https://arxiv.org/abs/1007.5424}{{\tt 1007.5424}}].

\bibitem{Padilla:2010ir}
A.~Padilla, P.~M. Saffin and S.-Y. Zhou, \emph{{Multi-galileons, solitons and
  Derrick's theorem}},
  \href{http://dx.doi.org/10.1103/PhysRevD.83.045009}{\emph{Phys. Rev. D} {\bf
  83} (2011) 045009}, [\href{https://arxiv.org/abs/1008.0745}{{\tt
  1008.0745}}].

\bibitem{Hinterbichler:2010xn}
K.~Hinterbichler, M.~Trodden and D.~Wesley, \emph{{Multi-field galileons and
  higher co-dimension branes}},
  \href{http://dx.doi.org/10.1103/PhysRevD.82.124018}{\emph{Phys. Rev. D} {\bf
  82} (2010) 124018}, [\href{https://arxiv.org/abs/1008.1305}{{\tt
  1008.1305}}].

\bibitem{Bakhmatov:2019dow}
I.~Bakhmatov, N.~S. Deger, E.~T. Musaev, E.~O. Colg\'ain and M.~M.
  Sheikh-Jabbari, \emph{{Tri-vector deformations in $d=11$ supergravity}},
  \href{http://dx.doi.org/10.1007/JHEP08(2019)126}{\emph{JHEP} {\bf 08} (2019)
  126}, [\href{https://arxiv.org/abs/1906.09052}{{\tt 1906.09052}}].

\bibitem{Kobayashi:2019hrl}
T.~Kobayashi, \emph{{Horndeski theory and beyond: a review}},
  \href{http://dx.doi.org/10.1088/1361-6633/ab2429}{\emph{Rept. Prog. Phys.}
  {\bf 82} (2019) 086901}, [\href{https://arxiv.org/abs/1901.07183}{{\tt
  1901.07183}}].

\bibitem{Zumalacarregui:2013pma}
M.~Zumalac\'arregui and J.~Garc\'\i{}a-Bellido, \emph{{Transforming gravity:
  from derivative couplings to matter to second-order scalar-tensor theories
  beyond the Horndeski Lagrangian}},
  \href{http://dx.doi.org/10.1103/PhysRevD.89.064046}{\emph{Phys. Rev. D} {\bf
  89} (2014) 064046}, [\href{https://arxiv.org/abs/1308.4685}{{\tt
  1308.4685}}].

\bibitem{Bettoni:2013diz}
D.~Bettoni and S.~Liberati, \emph{{Disformal invariance of second order
  scalar-tensor theories: Framing the Horndeski action}},
  \href{http://dx.doi.org/10.1103/PhysRevD.88.084020}{\emph{Phys. Rev. D} {\bf
  88} (2013) 084020}, [\href{https://arxiv.org/abs/1306.6724}{{\tt
  1306.6724}}].

\bibitem{Chamseddine:2013kea}
A.~H. Chamseddine and V.~Mukhanov, \emph{{Mimetic Dark Matter}},
  \href{http://dx.doi.org/10.1007/JHEP11(2013)135}{\emph{JHEP} {\bf 11} (2013)
  135}, [\href{https://arxiv.org/abs/1308.5410}{{\tt 1308.5410}}].

\bibitem{Langlois:2018jdg}
D.~Langlois, M.~Mancarella, K.~Noui and F.~Vernizzi, \emph{{Mimetic gravity as
  DHOST theories}},
  \href{http://dx.doi.org/10.1088/1475-7516/2019/02/036}{\emph{JCAP} {\bf 02}
  (2019) 036}, [\href{https://arxiv.org/abs/1802.03394}{{\tt 1802.03394}}].

\bibitem{Crisostomi:2017ugk}
M.~Crisostomi, K.~Noui, C.~Charmousis and D.~Langlois, \emph{{Beyond Lovelock
  gravity: Higher derivative metric theories}},
  \href{http://dx.doi.org/10.1103/PhysRevD.97.044034}{\emph{Phys. Rev. D} {\bf
  97} (2018) 044034}, [\href{https://arxiv.org/abs/1710.04531}{{\tt
  1710.04531}}].

\bibitem{sveskomantoneassontba}
E.~Easson, T.~Manton and A.~Svesko, \emph{{To Appear}}.

\bibitem{Buchdahl:1979wi}
H.~A. Buchdahl, \emph{{On a Lagrangian for Nonminimally Coupled Gravitational
  and electromagnetic fields}},
  \href{http://dx.doi.org/10.1088/0305-4470/12/7/020}{\emph{J. Phys. A} {\bf
  12} (1979) 1037--1043}.

\bibitem{Horndeski:1976gi}
G.~Horndeski, \emph{{Conservation of Charge and the Einstein-Maxwell Field
  Equations}}, \href{http://dx.doi.org/10.1063/1.522837}{\emph{J. Math. Phys.}
  {\bf 17} (1976) 1980--1987}.

\bibitem{Taylor:2008tg}
M.~Taylor, \emph{{Non-relativistic holography}},
  \href{https://arxiv.org/abs/0812.0530}{{\tt 0812.0530}}.

\bibitem{Charmousis:2010zz}
C.~Charmousis, B.~Gouteraux, B.~S. Kim, E.~Kiritsis and R.~Meyer,
  \emph{{Effective Holographic Theories for low-temperature condensed matter
  systems}}, \href{http://dx.doi.org/10.1007/JHEP11(2010)151}{\emph{JHEP} {\bf
  11} (2010) 151}, [\href{https://arxiv.org/abs/1005.4690}{{\tt 1005.4690}}].

\bibitem{Lee:2010ii}
B.-H. Lee, D.-W. Pang and C.~Park, \emph{{Strange Metallic Behavior in
  Anisotropic Background}},
  \href{http://dx.doi.org/10.1007/JHEP07(2010)057}{\emph{JHEP} {\bf 07} (2010)
  057}, [\href{https://arxiv.org/abs/1006.1719}{{\tt 1006.1719}}].

\bibitem{Liu:2010ka}
Y.~Liu and Y.-W. Sun, \emph{{Holographic Superconductors from
  Einstein-Maxwell-Dilaton Gravity}},
  \href{http://dx.doi.org/10.1007/JHEP07(2010)099}{\emph{JHEP} {\bf 07} (2010)
  099}, [\href{https://arxiv.org/abs/1006.2726}{{\tt 1006.2726}}].

\bibitem{Gath:2012pg}
J.~Gath, J.~Hartong, R.~Monteiro and N.~A. Obers, \emph{{Holographic Models for
  Theories with Hyperscaling Violation}},
  \href{http://dx.doi.org/10.1007/JHEP04(2013)159}{\emph{JHEP} {\bf 04} (2013)
  159}, [\href{https://arxiv.org/abs/1212.3263}{{\tt 1212.3263}}].

\bibitem{OKeeffe:2013xdv}
D.~K. O'Keeffe and A.~W. Peet, \emph{{Electric hyperscaling violating solutions
  in Einstein-Maxwell-dilaton gravity with $R^2$ corrections}},
  \href{http://dx.doi.org/10.1103/PhysRevD.90.026004}{\emph{Phys. Rev. D} {\bf
  90} (2014) 026004}, [\href{https://arxiv.org/abs/1312.2261}{{\tt
  1312.2261}}].

\bibitem{Li:2016rcv}
L.~Li, \emph{{Hyperscaling Violating Solutions in Generalised EMD Theory}},
  \href{http://dx.doi.org/10.1016/j.physletb.2017.02.004}{\emph{Phys. Lett. B}
  {\bf 767} (2017) 278--284}, [\href{https://arxiv.org/abs/1608.03247}{{\tt
  1608.03247}}].

\bibitem{Pedraza:2018eey}
J.~F. Pedraza, W.~Sybesma and M.~R. Visser, \emph{{Hyperscaling violating black
  holes with spherical and hyperbolic horizons}},
  \href{http://dx.doi.org/10.1088/1361-6382/ab0094}{\emph{Class. Quant. Grav.}
  {\bf 36} (2019) 054002}, [\href{https://arxiv.org/abs/1807.09770}{{\tt
  1807.09770}}].

\bibitem{Gouteraux:2011ce}
B.~Gouteraux and E.~Kiritsis, \emph{{Generalized Holographic Quantum
  Criticality at Finite Density}},
  \href{http://dx.doi.org/10.1007/JHEP12(2011)036}{\emph{JHEP} {\bf 12} (2011)
  036}, [\href{https://arxiv.org/abs/1107.2116}{{\tt 1107.2116}}].

\bibitem{Elvang:2017mdq}
H.~Elvang, M.~Hadjiantonis, C.~R. Jones and S.~Paranjape, \emph{{On the
  Supersymmetrization of Galileon Theories in Four Dimensions}},
  \href{http://dx.doi.org/10.1016/j.physletb.2018.04.032}{\emph{Phys. Lett. B}
  {\bf 781} (2018) 656--663}, [\href{https://arxiv.org/abs/1712.09937}{{\tt
  1712.09937}}].

\bibitem{Padmanabhan:2007en}
T.~Padmanabhan and A.~Paranjape, \emph{{Entropy of null surfaces and dynamics
  of spacetime}},
  \href{http://dx.doi.org/10.1103/PhysRevD.75.064004}{\emph{Phys. Rev. D} {\bf
  75} (2007) 064004}, [\href{https://arxiv.org/abs/gr-qc/0701003}{{\tt
  gr-qc/0701003}}].

\bibitem{Svesko:2020yxo}
A.~Svesko, \emph{{Emergence of Spacetime: From Entanglement to Einstein}},
  \href{https://arxiv.org/abs/2006.13106}{{\tt 2006.13106}}.

\end{thebibliography}\endgroup

\end{document}